\documentclass[final,5p,twocolumn]{elsarticle}

\usepackage{graphicx}

\usepackage{amssymb}

\usepackage{latexsym}
\usepackage[fleqn]{amsmath}
\usepackage{amssymb}
\usepackage{float}
\usepackage{fancyvrb}
\fvset{fontsize=\small}
\usepackage{listings}
\newcounter{bla}

\journal{Computer Physics Communications}

\begin{document}

\begin{frontmatter}



\title{ZKCM: a C++ library for multiprecision matrix computation with applications in quantum information}


\author[a]{Akira SaiToh\corref{author}}

\cortext[author] {Corresponding author.\\\textit{E-mail address:} akirasaitoh@nii.ac.jp}
\address[a]{Quantum Information Science Theory Group,
National Institute of Informatics,
2-1-2 Hitotsubashi, Chiyoda, Tokyo 101-8430, Japan}
\begin{abstract}
ZKCM is a C++ library developed for the purpose of multiprecision matrix computation,
on the basis of the GNU MP and MPFR libraries. It provides an easy-to-use syntax and
convenient functions for matrix manipulations including those often used in numerical
simulations in quantum physics. Its extension library, ZKCM\_QC, is developed for
simulating quantum computing using the time-dependent matrix-product-state simulation
method. This paper gives an introduction about the libraries with practical sample programs.
\end{abstract}


\end{frontmatter}



{\bf PROGRAM SUMMARY}

\begin{small}
\noindent
{\em Manuscript Title:} ZKCM: a C++ library for multiprecision
matrix computation with applications in quantum information\\
{\em Authors:} Akira SaiToh\\
{\em Program Title:} ZKCM\\
{\em Journal Reference:}                                      \\
{\em Catalogue identifier:}                                   \\
{\em Licensing provisions:} GNU Lesser General Public License Ver. 3\\
{\em Programming language:} C++\\
{\em Computer:} General computers\\
{\em Operating system:} Unix-like systems, such as Linux, Free BSD, Cygwin on Windows OS, etc.\\
{\em RAM:} Several mega bytes - several giga bytes, dependent on the problem instance
{\em Classification:} 4.8 Linear Equations and Matrices, 4.15 Quantum Computing\\
{\em External routines/libraries:} GNU MP (GMP) \cite{1}, MPFR \cite{2} Ver. 3.0.0 or later\\
{\em Nature of problem:}
Multiprecision computation is helpful to guarantee and/or evaluate the accuracy
of simulation results in numerical physics. There is a potential demand for
a programming library focusing on matrix computation usable for this purpose
with a user-friendly syntax.\\
{\em Solution method:}
A C++ library ZKCM has been developed for multiprecision matrix computation.
It provides matrix operations useful for numerical studies of physics, e.g., the
tensor product (Kronecker product), the tracing-out operation, the inner product,
the LU decomposition, the Hermitian-matrix diagonalization, the singular-value decomposition,
and the discrete Fourier transform. For basic floating-point operations, GMP and MPFR
libraries are used. An extension library ZKCM\_QC has also been developed,
which employs the time-dependent matrix-product-state method to simulate
quantum computing.\\
{\em Restrictions:}
Multiprecision computation with more than a half thousand bit precision
is often a thousand times slower than double-precision computation for any
kind of matrix computation.
   \\
{\em Additional comments:}
A user's manual is placed in the directory ``doc'' of the package.
Each function is explained in a reference manual found in the directories
``doc/html'' and ``doc/latex''. Sample programs are placed in the directory
``samples''.\\
{\em Running time:}
It takes less than thirty seconds to obtain a DFT spectrum for
$2^{16}$ data points of a time evolution of a quantum system described
by a $4\times4$ matrix Hamiltonian for 256-bit precision when we use recent AMD or
Intel CPU with 2.5 GHz or more CPU frequency. It takes three to five minutes to
diagonalize a $100\times100$ Hermitian matrix for 512-bit precision using the
aforementioned CPU.\\

\end{small}


\section{Introduction}
Precision of floating-point operations is sometimes of serious concern in
simulation physics when rounding errors in variables or matrix elements
considerably affect numerical results for investigated physical phenomena
 \cite{BBB12}. There are several programming libraries, e.g.,
Refs.~\cite{FMZM,CLN,ARPREC,Exflib,MPACK,ZKCM}, for high-precision computing, which
are helpful in this regard. Among them, the library named ZKCM \cite{ZKCM},
which we have been developing, is a C++ library for multiprecision
complex-number matrix computation. It provides several functionalities
including the LU decomposition, the singular value decomposition, the tensor
product, and the tracing-out operation and an easy-to-use syntax
for basic operations. It is based on the GNU MP (GMP) \cite{GMP} and
MPFR \cite{MPFR} libraries, which are commonly included in recent
distributions of UNIX-like systems.

There is an extension library named ZKCM\_QC. This library is 
designed for simulating quantum computing \cite{Gruska,NC2000} by the
time-dependent matrix-product-state method \cite{V03} [or, simply referred
to as the matrix-product-state (MPS) method]. It uses a matrix product state
\cite{WH92,WH93,V03} to represent a pure quantum state.
The MPS method is recently one of the standard methods for
simulation-physics software \cite{ALPS}. As for other methods effective
for simulating quantum computing, see, e.g., Refs.\ \cite{VMH03,AG04}.
With ZKCM\_QC, one may use quantum gates in $\rm{U(2)}$, ${\rm U(4)}$,
and ${\rm U(8)}$ as elementary gates. Indeed, in general, quantum
gates in ${\rm U(2)}$ and ${\rm U(4)}$ are enough for universal quantum
computing \cite{DBE95}, but we regard quantum gates in ${\rm U(8)}$ also as
elementary gates so as to reduce computational overheads in circuit
constructions.

A simulation of quantum computing with MPS is known for its
computational efficiency in case the Schmidt ranks are kept small
during the simulation \cite{V03,KW04}.
Even for the case slightly large Schmidt ranks are involved, it is
not as expensive as a simple simulation. The theory of MPS simulation
will be briefly explained in Sec.\ \ref{sectheory}. Numerical errors will be
phenomenologically discussed in Sec.\ \ref{secDJ}, which will give a certain
reason why we introduce multiprecision computation for an MPS simulation
of quantum computing. 

This contribution is intended to provide a useful introduction
for programming with the libraries. Section\ \ref{secZKCM} describes
two examples of the use of the ZKCM library: one for showing the precision
dependence of a solution of a simple linear equation; the other for
simulating an NMR spectrum in a simple model. Performance evaluation of
the library is made in Sec.\ \ref{secPE}. Section\ \ref{secZKCMQC}
shows an overview of the theory of the MPS method and an example of
simulating a simple quantum circuit using the ZKCM\_QC library.
In addition, later in the section numerical errors in an MPS simulation of
quantum computing are examined using a certain setup of a quantum algorithm.
We discuss on the effectiveness and the performance of our libraries in
Sec.\ \ref{secDiscussion}. Section \ref{secSummary} summarizes our
achievements.

\section{ZKCM Library}\label{secZKCM}
The ZKCM library is designed for general-purpose matrix computation.
This section concentrates on its main library. It has two major
C++ classes: ``\verb|zkcm_class|'' and ``\verb|zkcm_matrix|''. The former
class is a class of a complex number. Many operators like ``\verb|+=|'' and
functions like trigonometric functions are defined for the class. The latter
class is a class of a matrix. Standard operations and functions like matrix
inversion are defined. In addition, the singular-value decomposition of a
general matrix, the diagonalization of an Hermitian matrix, 
the discrete Fourier transform, etc., are defined for the class.
Functoins for the tensor product (Kronecker product) and
the tracing-out operation (e.g., one can trace out the subsystem B of
system ABC) are also defined.
A detailed document is placed in the ``doc'' directory of the
package of ZKCM.

As for installation of the library, the standard process
``\verb|./configure| $\rightarrow$ \verb|make| $\rightarrow$ \verb|sudo make install|''
works in most cases; otherwise the document should be consulted.

We will next look at simple examples to demonstrate the programming
style using the library. (Here, the latest stable version, ver. 0.3.2, is used.)

\subsection{Program examples}
\paragraph{Example 1}
There is a classical example \cite{KM86} often mentioned to
recall the importance of multiprecision computation. It clarifies
that a sufficiently large precision is required even for a simple
linear equation with only two variables.

Consider the linear equation
\[
A \begin{pmatrix}x\\y\end{pmatrix}
=\begin{pmatrix}1\\0\end{pmatrix}
\]
with
\[
A= \begin{pmatrix}
64919121 & -159018721\\
41869520.5 & -102558961
\end{pmatrix}.
\]
The exact solution is $x=205117922, y=83739041$.

\begin{figure}[H]
\begin{center}
\includegraphics[width=79mm]{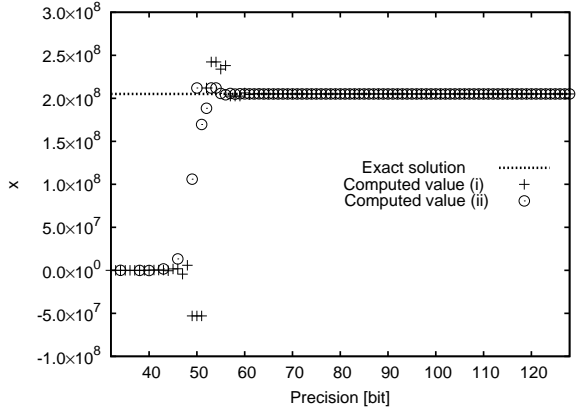}
\caption{\label{figclassical}Plot of the computed values of $x$ against
the precision (the bit length of a mantissa) employed for floating-point numbers.
Either (i) the matrix inversion (pluses) or (ii) the Gaussian elimination (circles) was used.
For case (ii), there are data points with $x$ values $< -1.0\times 10^8$ for some values
of precision $<50$; they are outside the range in the figure.}
\end{center}
\end{figure}

One way to numerically find the solution is to compute $(x,y)^t = A^{-1}(1,0)^t$
using the matrix inversion. Another way is to utilize the Gaussian elimination (see,
{\em e.g.}, Refs.~\cite{W61,T85}).
We used functions ``\verb|inv|'' and ``\verb|gauss|'' of ZKCM for the former and
the latter ways, respectively (as for pivoting, a partial pivoting is used in ``\verb|gauss|'').
Then, we plotted the computed value of $x$ against the precision [bits] as shown in
Fig.\ \ref{figclassical}. (Here, we refer to the number of bits for a significand
of a floating-point number as precision.\footnote{Precision is defined as an exact length of
the mantissa portion of a floating-point number in MPFR while it is defined as a least length
in GMP \cite{MPFR}. In ZKCM, a complex number keeps each part internally as an MPFR variable
so that precision is an exact length. However, it should be noted that an output from a
function usually carries over the best precision among those of involved variables in ZKCM.})
We can see that low-precision computing resulted in a wrong answer while
high-precision computing resulted in a correct answer. It should be
emphasized that the double precision (namely, 53 bits for the mantissa
portion of a floating-point number) is not enough in this example.
The program used for this example is shown in Listing~\ref{codeclassic}.
It is found as ``samples/classic\_example.cpp'' in the package. It is written
to use the matrix inversion. To use the Gaussian elimination, the line
``\verb|B = inv(A) * B;|'' should be replaced with
``\verb|B = gauss(A, B);|''.
\begin{lstlisting}[label=codeclassic,caption=classic\_example.cpp]
#include "zkcm.hpp"
#include <fstream>
int main()
{
  std::ofstream ofs("classic_example.dat");
  if (!ofs)
  {
    std::cerr << "Could not create the file\
 classic_example.dat"
              << std::endl;
    return -1;
  }
  ofs << "# prec x y" << std::endl;
  for (int prec = 32; prec < 257; prec++)
  {
    zkcm_set_default_prec(prec);
    zkcm_matrix A(2, 2), B(2, 1);

    A(0, 0) = "64919121";
    A(0, 1) = "-159018721";
    A(1, 0) = "41869520.5";
    A(1, 1) = "-102558961";

    B(0, 0) = 1;
    B(1, 0) = 0;

    B = inv(A) * B;
    //Replace the above line with
    //B = gauss(A, B);
    //to use the Gaussian elimination instead.

    ofs << prec << " "
        << B(0, 0) << " "
        << B(1, 0) << std::endl;
  }
  zkcm_quit();
  return 0;
}
\end{lstlisting}
The program is compiled and executed in a standard way.\footnote{To
make an executable file, the library flags typically
``-lzkcm -lm -lmpfr -lgmp -lgmpxx'' are required. As for
ZKCM\_QC, additionally ``-lzkcm\_qc'' should be specified.}
Its output file ``classic\_example.dat'' can be visualized 
by Gnuplot \cite{Gnuplot} using the file ``classic\_example.gnuplot'' found
in the ``samples'' directory.

In the program, one can see typical features of the ZKCM library.
The line ``\verb|zkcm_set_default_prec(prec);|'' sets the default internal
precision (the default bit length of a significand) of an object to ``\verb|prec|''.
Any object like ``\verb|A|'' (this is a $2\times2$ matrix) constructed
without specified precision possesses the default precision. It is possible
to specify a particular precision ({\em i.e.}, in case of a matrix object,
the length of the significand for each part of each element of the matrix),
say, 512, by constructing the object as ``\verb|zkcm_matrix A(2, 2, 512)|'', for
instance.
For a matrix object, say, ``\verb|A|'', it is convenient to access its
$(i,j)$ element by ``\verb|A(i,j)|'' that is a reference to the element.
The element is an object in the type of ``\verb|zkcm_class|''.
To assign a value into a ``\verb|zkcm_class|'' object (a complex number) ``\verb|z|'',
one can write ``\verb|z=...;|'' where the right-hand side can be a number or
a string describing a complex number in the style \verb|"___+___*I"|
(PARI/GP style \cite{PARI}; here, ``\verb|I|'' stands for $i=\sqrt{-1}$),
\verb|"___+___i"|, \verb|"___+___j"| (here, ``\verb|j|'' stands for $i$), etc.
(there are other acceptable styles).
In the program, by ``\verb|A(0, 0) = "64919121";|'' the value $64919121+0i$ is
assigned to the $(0,0)$ element of matrix \verb|A|. Other values are assigned
to corresponding elements in a similar way. The solution of the linear equation
is obtained by using the function ``\verb|inv|'' or ``\verb|gauss|''.
The obtained result is written into the output file stream ``\verb|ofs|'' by
using the operator ``\verb|<<|''. In the program, the elements are individually
written out so as to meet the data format of a data-plotting program.
After computation is performed, the program should be terminated. At this stage,
it is recommended to write ``\verb|zkcm_quit();|'' so as to release miscellaneous
memories allocated for internal use of background library functions.

\paragraph{Example 2}
As the second example, a sample program ``NMR\_spectrum\_simulation.cpp''
found in the ``samples'' directory of the package of ZKCM
is explained. This program generates a simulated free-induction-decay (FID)
spectrum of liquid-state NMR for the spin system consisting of a
proton spin with precession frequency $w_1=$~400 MHz (variable ``\verb|w1|'' in
the program) and a ${}^{13}{\rm C}$ spin with precession frequency $w_2=$~125 MHz
(variable ``\verb|w2|'') at room temperature (300 K) (variable ``\verb|T|'').
A J coupling constant $J_{12}=140$ kHz (variable ``\verb|J12|'') is considered
for the spins.

The first line of the program is to include a header file of ZKCM:
\begin{Verbatim}
#include "zkcm.hpp"
int main(int argc, char *argv[])
{
\end{Verbatim}
In the beginning of the ``\verb|main|'' function, the default precision is
set to 280 bits by
\begin{Verbatim}
  zkcm_set_default_prec(280);
\end{Verbatim}
In the subsequent lines, some matrices like Pauli matrices $I$ and $X$, etc.
are generated. For example, $X$ is generated as
\begin{Verbatim}
  zkcm_matrix X = "[0, 1; 1, 0]";
\end{Verbatim}
using a string representing a matrix in the PARI/GP style.
The $Y_{90}$ pulse is generated as
\begin{Verbatim}
  zkcm_matrix Yhpi(2,2);
  Yhpi(0,0) = sqrt(zkcm_class(0.5));
  Yhpi(0,1) = sqrt(zkcm_class(0.5));
  Yhpi(1,0) = -sqrt(zkcm_class(0.5));
  Yhpi(1,1) = sqrt(zkcm_class(0.5));  
\end{Verbatim}
Other matrices are generated by similar lines.
After this, values of constants and parameters are set.
For example, the Boltzmann constant $k_{\rm B}$ [J/K] is generated as
\begin{Verbatim}
  zkcm_class kB("1.3806504e-23");
\end{Verbatim}
The Hamiltonian $H$ of the spin system is generated in the type of
``\verb|zkcm_matrix|'' and is specified as
\begin{Verbatim}
  H = w1 * tensorprod(Z/2,I) + w2 * tensorprod(I,Z/2)
      + J12 * tensorprod(Z/2,Z/2);
\end{Verbatim}
This is used to generate a thermal state $\rho$:
\begin{Verbatim}
  zkcm_matrix rho(4,4);
  rho = exp_H((-hplanck/kB/T) * H);
  rho /= trace(rho);
\end{Verbatim}
Here, ``\verb|exp_H|'' is a function to calculate the exponential of an
Hermitian matrix and ``\verb|hplanck|'' is the Planck constant
($6.62606896\times 10^{-34}$ Js).
The sampling time interval $dt$ to record the value of $<X>$ for the proton
spin is set to $0.43/w_1$ (any number smaller than $1/2$ might be fine instead
of $0.43$) by
\begin{Verbatim}
  zkcm_class dt(zkcm_class("0.43")/w1);
\end{Verbatim}
The number of data to record is then decided as
\begin{Verbatim}
  int N = UNP2(8.0/dt/J12);
\end{Verbatim}
Here, function ``\verb|UNP2|'' returns the integer upper nearest power of 2 for a
given number.
Now arrays to store data are prepared as row vectors.
\begin{Verbatim}
  zkcm_matrix array(1, N), array2(1, N);
\end{Verbatim}
The following lines prepare the $X$ and $Y_{90}$-pulse operators
acting only on the proton spin.
\begin{Verbatim}
  zkcm_matrix X1(4, 4), Yhpi1(4, 4);
  X1 = tensorprod(X, I);
  Yhpi1 = tensorprod(Yhpi, I);
\end{Verbatim}
To get an FID data, we firstly tilt the proton spin by the ideal $Y_{90}$ pulse.
\begin{Verbatim}
  rho = Yhpi1 * rho * adjoint(Yhpi1);
\end{Verbatim}
Now the data of time evolution of $<X>$ of the proton spin under the Hamiltonian
$H$ is recorded for the time duration ${\rm N} \times {\rm dt}$ using
\begin{Verbatim}
  array = rec_evol(rho, H, X1, dt, N);
\end{Verbatim}
We now use a zero-padding for this ``\verb|array|'' so as to enhance the resolution.
This will extend the array by N zeros. 
\begin{Verbatim}
  array2 = zero_padding(array, 2*N);
\end{Verbatim}
To obtain a spectrum, the discrete Fourier transform is applied.
\begin{Verbatim}
  array2 = abs(DFT(array2));
\end{Verbatim}
This ``\verb|array2|'' is output to the file ``example\_zp.fid'' as an FID spectrum
data with $df = 1/(2\times N\times dt)$ as the frequency interval, in the
Gnuplot style by
\begin{Verbatim}
  GP_1D_print(array2, 1.0/dt/zkcm_class(2*N),
              1, "example_zp.fid");
\end{Verbatim}
At last, the function ``\verb|main|'' ends with ``\verb|zkcm_quit();|'' and 
``\verb|return 0;|''.
The program is compiled and executed in a standard
way.

\begin{figure}[H]
\begin{center}
\includegraphics[width=79mm]{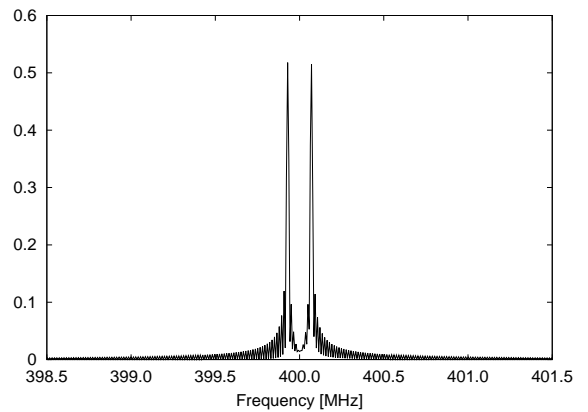}
\end{center}
\caption{\label{figzp} Plot of the simulation data stored in ``example\_zp.fid''.
See the text for the details of this simulation.}
\end{figure}

The result stored in ``example\_zp.fid'' is visualized by Gnuplot using 
the file ``example\_zp.gnuplot'' placed in the directory ``samples'', as
shown in Fig.\ \ref{figzp}.
It should be noted that we have not employed a high-temperature
approximation. Under a high-temperature approximation, the first order
deviation $-\beta H$ of $\exp(-\beta H)$ (here, $\beta=h/(k_{\rm B}T)$)
is considered as a deviation density matrix and calculations are
performed using the normalized deviation density matrix $-H/{\rm const.}$
This approximation is commonly used \cite{GAMMA} but it cannot be
used for simulations for low temperature.
An advantage of using ZKCM for simulating NMR spectra
is that the temperature can be chosen. This is possible because of
high accuracy in computing the exponential of a Hamiltonian.

As we have seen in this example, ZKCM has useful functions to handle
matrices, especially those often used in simulations in quantum physics.
For the list of available functions, see the reference manual found
in the package.

\section{Performance evaluation}\label{secPE}
We evaluate the performance of ZKCM (version 0.3.2) in comparison
with PARI \cite{PARI} (version 2.5.3 with the GMP kernel),
a conventional highly-evaluated C library for mathematical computing.
\subsection{Speed of computing the $\Gamma$ function}
Here, we evaluate the performance of the function ``\verb|gamma|'' of ZKCM,
which calculates $\Gamma(z)$ for a complex number $z$.
It is implemented on the basis of the expansion using the Stirling
formula for ${\rm Re}\;z \ge 0$:
\[
\Gamma(z)\simeq e^{-z}z^{z-1}\sqrt{2\pi z}\exp\left[
\sum_{n=1}^{L}\frac{(-1)^{n-1}|B_{2n}|z^{1-2n}}{2n(2n-1)}+R_L
\right],
\]
where $B_m$ is the first or the second Bernoulli number (as the absolute
value is used, either is fine); $L$ is a sufficiently large (but not very large)
integer to guarantee the accuracy; $R_L$ is the remainder of the sum (corresponding
to the sum of the terms for $n=L+1,\ldots,\infty$). It is known that the expansion
is not convergent since $|B_{2n}|$ grows rapidly for large $n$. It was, however,
proved by Spira \cite{Sp71} that $|R_L|\le \frac{|B_{2L}|}{2L-1}|z|^{1-2L}$ holds
when ${\rm Re}\,z\ge 0$. Thus the expansion is usable for computation with enough
accuracy as long as $z$ is appropriately scaled to a large value in advance.
We internally scale $z$ to satisfy ${\rm Re}\,z\ge 0$ and $|z|\ge 10^3\sim 2^{10}$.
In accordance with this scaling, $L$ is appropriately chosen so that $R_L$ becomes
small enough for required precision. For the scaling, the well-known formulae
$\Gamma(z) = \Gamma(z+k) / [z(z+1)\cdots(z+k-1)]$ (here, $k$ is an integer)
and
$\Gamma(z) = -\pi / [-z\Gamma(-z)\sin(-\pi z)]$ (here, $z$ is not a pole)
are utilized.
To compute Bernoulli numbers, we utilize the relation
\[
B_m = 1-\sum_{k=0}^{n-1}\left(\begin{array}{c}n\\k\end{array}\right)\frac{B_k}{n-k+1}
\]
and the fact that $B_{2j+1}$ vanishes for integer $j\ge 1$. We keep the computed values
of binomial coefficients and Bernoulli numbers inside a certain subblock of a function for
computing $\Gamma(z)$. Therefore, the cost to compute $B_m$ assuming that we have already
computed $B_0,\ldots,B_{m-2}$ for even number $m$ is small: $O(m)$ rational-number operations
are enough\footnote{
For other use than computing $\Gamma(z)$, we cannot use this assumption in general.
In this case, Akiyama-Tanigawa algorithm \cite{Kaneko00} is employed to calculate the Bernoulli
number. This takes $O(m^2)$ rational-number operations.} (we make use of the C++ class
``\verb|mpq_class|'' of GMP for internally handling rational numbers).

Now, the speed of computing $\Gamma(z)$ using ``\verb|gamma|'' is evaluated for $z=a+bi$
with random $a,b\in[-10,10]$. It is compared with the function with the same name,
``\verb|gamma|'', of the PARI library. The PARI library uses a similar algorithm to
compute $\Gamma(z)$. The main difference is that PARI can utilize cached values of
Bernoulli numbers that are once calculated in any function during a process is running.
It can also use precomputed values of typical constants.

As shown in Table\ \ref{tablegamma}, the average computation time of $\Gamma(z)$ in
the ZKCM library is on the same order of magnitude as that in the PARI library in the
absence of cached values of Bernoulli numbers. However, in the presence of cached
values of Bernoulli numbers, the computation speed of $\Gamma(z)$ in PARI is quite
much faster.

\begin{table}[pbt]
\caption{\label{tablegamma}Comparison of the average real time consumption for
computing $\Gamma(z)$ for $z=a+bi$ with random $a,b\in[-10,10]$. The average was
taken over 1,000 trials. No I/O operation was involved. Time for generating $z$
was negligible (this was no more than $2.0\times 10^{-5}$ seconds). The standard
deviation (we employ the sample standard deviation) for each entry is shown in parentheses
in small fonts under the entry.
``prec'' stands for the precision (the bit length of each mantissa) employed in a
program. In a program using PARI, function ``{\tt nbits2prec}'' was used to convert the
precision into the number of code words that is defined as the precision for variables
in PARI. ZKCM version 0.3.2 and PARI version 2.5.3 were used. They
were compiled with GCC/G++ with their default optimization flags, while test programs
for this performance evaluation were compiled without optimization flags.\protect\footnotemark[4]
In this table, the middle and the right columns show the results for PARI with and without
precomputed values of Bernoulli numbers, respectively. Environments: For upper entries
($\blacklozenge$): Fedora 15 64-bit operating system on Intel Core i5 M460 CPU
(2.53 GHz, Max. 2.80GHz) with 4GB RAM. For lower entries ($\clubsuit$): Fedora 16 64-bit operating
system on AMD FX-8120 CPU (3.10GHz, Max. 4.00GHz) with 16GB RAM.}
\begin{center}
\begin{tabular}{|r|p{2cm}|p{2cm}|p{2cm}|}\hline
prec&ZKCM [sec]& PARI without cache [sec]&PARI with cache [sec]\\\hline
 256 & $\begin{array}{l}{\hspace{-1.8mm}{}^\blacklozenge}3.73\times10^{-3}\\\text{\scriptsize($8.72\times10^{-5}$)}\\{\hspace{-2mm}{}^\clubsuit}2.95\times10^{-3}\\\text{\scriptsize($9.58\times10^{-5}$)}\end{array}$ & $\begin{array}{l}1.38\times10^{-3}\\\text{\scriptsize{($7.91\times10^{-5}$)}}\\1.42\times10^{-3}\\\text{\scriptsize($2.73\times10^{-4}$)}\end{array}$ & $\begin{array}{l}1.63\times10^{-4}\\\text{\scriptsize($6.10\times10^{-5}$)}\\1.48\times10^{-4}\\\text{\scriptsize($9.56\times10^{-5}$)}\end{array}$\\\hline
 512 & $\begin{array}{l}9.04\times10^{-3}\\\text{\scriptsize($2.95\times10^{-4}$)}\\7.15\times10^{-3}\\\text{\scriptsize($1.64\times10^{-4}$)}\end{array}$ & $\begin{array}{l}4.70\times10^{-3}\\\text{\scriptsize($6.79\times10^{-4}$)}\\3.12\times10^{-3}\\\text{\scriptsize($6.78\times10^{-4}$)}\end{array}$ & $\begin{array}{l}3.43\times10^{-4}\\\text{\scriptsize($1.49\times10^{-4}$)}\\3.43\times10^{-4}\\\text{\scriptsize($1.75\times10^{-4}$)}\end{array}$\\\hline
 768 & $\begin{array}{l}1.86\times10^{-2}\\\text{\scriptsize($2.62\times10^{-4}$)}\\1.45\times10^{-2}\\\text{\scriptsize($2.80\times10^{-4}$)}\end{array}$ & $\begin{array}{l}9.58\times10^{-3}\\\text{\scriptsize($1.82\times10^{-3}$)}\\8.38\times10^{-3}\\\text{\scriptsize($1.85\times10^{-3}$)}\end{array}$ & $\begin{array}{l}6.61\times10^{-4}\\\text{\scriptsize($3.67\times10^{-4}$)}\\6.02\times10^{-4}\\\text{\scriptsize($2.07\times10^{-4}$)}\end{array}$\\\hline
1024 & $\begin{array}{l}3.48\times10^{-2}\\\text{\scriptsize($8.77\times10^{-4}$)}\\2.59\times10^{-2}\\\text{\scriptsize($3.67\times10^{-4}$)}\end{array}$ & $\begin{array}{l}1.48\times10^{-2}\\\text{\scriptsize($8.32\times10^{-4}$)}\\1.34\times10^{-2}\\\text{\scriptsize($6.35\times10^{-4}$)}\end{array}$ & $\begin{array}{l}1.25\times10^{-3}\\\text{\scriptsize($3.39\times10^{-4}$)}\\1.03\times10^{-3}\\\text{\scriptsize($3.64\times10^{-4}$)}\end{array}$\\\hline
1280 & $\begin{array}{l}5.37\times10^{-2}\\\text{\scriptsize($1.41\times10^{-3}$)}\\3.92\times10^{-2}\\\text{\scriptsize($9.90\times10^{-4}$)}\end{array}$ & $\begin{array}{l}2.44\times10^{-2}\\\text{\scriptsize($9.68\times10^{-4}$)}\\2.09\times10^{-2}\\\text{\scriptsize($1.56\times10^{-3}$)}\end{array}$ & $\begin{array}{l}1.95\times10^{-3}\\\text{\scriptsize($6.87\times10^{-4}$)}\\1.70\times10^{-3}\\\text{\scriptsize($7.23\times10^{-4}$)}\end{array}$\\\hline
\end{tabular}
\end{center}
\end{table}
\footnotetext[4]{
This was because the improvement in running time due to setting compiler optimization flags was negligible.
(Of course, there were other compiler flags for library specifications etc. which are not of our concern in this context.)
For instance, a program to find eigenvectors of a normalized random $50 \times 50$
Hermitian matrix using the function ``{\tt diag\_H}'' of ZKCM (``{\tt eigen}'' of PARI) with
768-bit precision consumed 34.558 (14.550) seconds on average when compiled without optimization
flags, and consumed 34.314 (14.551) seconds on average when compiled with the optimization flags
``-O3 -fno-strict-aliasing -fomit-frame-pointer''. (Here, we used the Frobenius norm for normalization.)
The average was taken over 100 trials. The standard deviations were 0.0648 (0.143) and 0.0645 (0.122),
respectively. The compiler was GCC ver. 4.6.3 installed by default on the operating system Fedora 15
64-bit. A computer with the Intel Core i5 M460 2.53 GHz CPU (maximum clock frequency 2.80GHz) and 4GB RAM was used.}
\setcounter{footnote}{4}
\subsection{Speed of matrix multiplication}
Here, we evaluate the speed of matrix multiplication.
We first generate a $100\times 100$ random Hermitian matrix $A$ satisfying
$\sqrt{{\rm Tr}AA^\dagger}=1$. We then continue to compute the operations
(i) $A\leftarrow A^2$ and (ii) $A\leftarrow A/\sqrt{{\rm Tr}AA^\dagger}$
sequentially, {\em i.e.}, (i) (ii) (i) (ii) ... from left to right. Operations
are performed under a specified precision ``prec''. We compare time consumption
to perform ten sets of ``(i) (ii)''. Time spent for the matrix preparation is not
included. As shown in Table \ref{tableMM}, computation time is on the same order of
magnitude for ZKCM and PARI and discrepancy in time consumption is not
large. This result is plausible since both the libraries employ a straight-forward
way to implement matrix multiplication and rely on the GMP library for the speed
of primitive operations. (ZKCM uses MPFR for basic operations that is based on
GMP; hence ZKCM indirectly uses primitive operations of GMP.)
\begin{table}[pbt]
\caption{\label{tableMM} Comparison of the average real time consumption to perform
ten sets of operations (i) and (ii) (see the text). The average was taken over ten
different $A$'s (again, see the text). The standard deviation is shown in
parentheses in small fonts. ``prec'' stands for the precision (the bit length of a significand).
Environments: Fedora 15 OS on Intel Core i5 M460 CPU with 4GB RAM for upper entries ($\blacklozenge$)
and Fedora 16 OS on AMD FX-8120 CPU with 16GB RAM for lower entries ($\clubsuit$).}
\begin{center}  
\begin{tabular}{|r|c|c|}\hline
prec&ZKCM [sec]& PARI [sec]\\\hline
 256 & $\begin{array}{l}{\hspace{-1.8mm}{}^\blacklozenge}26.1\text{\scriptsize~(0.115)}\\{\hspace{-2mm}{}^\clubsuit}21.4\text{\scriptsize~(0.248)}\end{array}$ & $\begin{array}{l}15.0\text{\scriptsize~(0.204)}\\13.0\text{\scriptsize~(0.567)}\end{array}$ \\\hline
 512 & $\begin{array}{l}35.5\text{\scriptsize~(0.206)}\\27.3\text{\scriptsize~(0.202)}\end{array}$ & $\begin{array}{l}21.9\text{\scriptsize~(0.0208)}\\18.7\text{\scriptsize~(0.188)}\end{array}$ \\\hline
 768 & $\begin{array}{l}46.4\text{\scriptsize~(0.161)}\\36.9\text{\scriptsize~(0.266)}\end{array}$ & $\begin{array}{l}30.8\text{\scriptsize~(0.0702)}\\28.1\text{\scriptsize~(0.247)}\end{array}$ \\\hline
1024 & $\begin{array}{l}65.8\text{\scriptsize~(0.115)}\\53.2\text{\scriptsize~(0.178)}\end{array}$ & $\begin{array}{l}50.5\text{\scriptsize~(0.199)}\\41.3\text{\scriptsize~(0.434)}\end{array}$ \\\hline
1280 & $\begin{array}{l}86.5\text{\scriptsize~(0.0265)}\\69.3\text{\scriptsize~(0.181)}\end{array}$ & $\begin{array}{l}67.8\text{\scriptsize~(0.118)}\\54.9\text{\scriptsize~(0.715)}\end{array}$ \\\hline
\end{tabular}
\end{center}
\end{table}

\subsection{Speed of Hermitian-matrix diagonalization}
We next evaluate the speed of diagonalization of an Hermitian matrix
for finding all the eigenvectors.
The functions used for this task are
``\verb|diag_H|'' of ZKCM and ``\verb|eigen|'' and ``\verb|jacobi|'' of PARI.

Let us briefly explain our design of ``\verb|diag_H|''. In the function, the routine to
diagonalize an Hermitian matrix $A$ employs a standard QR method (see, {\em e.g.},
Refs.\ \cite{M66,K05}) to find eigenvalues, for which Householder reflections and Wilkinson shifts
are utilized. The eigenvalues $\lambda_i$ are sorted from larger to smaller.
Then each corresponding eigenvector $|v_i\rangle$ is calculated by the inverse iteration.
During this process, we set
\[A\leftarrow A+ {\rm const.}\times|v_{i-1}\rangle\langle v_{i-1}|\] before calculating $|v_i\rangle$.
This resolves degeneracy (in case we have), so that calculated eigenvector $|v_i\rangle$
becomes orthogonal to $|v_0\rangle, \ldots, |v_{i-1}\rangle$ with high accuracy. We still test
orthogonality and, in case required, perform an orthogonalization of $|v_i\rangle$'s.
Enhancement in accuracy of $\lambda_i$ is achieved during the inverse iteration steps. The
steps are retried, often with an initial vector set to the vector found in the previous round of steps,
and sometimes with a randomly chosen initial vector, until sufficient convergence of $|v_i\rangle$
and $\lambda_i$ for required precision is reached.

As for the functions of PARI, ``\verb|eigen|'' firstly computes the roots of the characteristic
polynomial and secondly uses the Gaussian elimination to find eigenvectors; in contrast,
``\verb|jacobi|'' simply uses the Jacobi method.
There are known facts on PARI library functions \cite{pariknown}: function ``\verb|eigen|'' uses
a naive algorithm so that it fails to compute eigenvectors for matrices with degenerate
eigenvalues; function ``\verb|jacobi|'' handles real symmetric matrices only, so that an Hermitian
matrix $A$ should be reformed into a symmetric matrix
$\begin{pmatrix}{\rm Re}(A) & -{\rm Im}(A)\\{\rm Im}(A) & {\rm Re}(A)\end{pmatrix}$
(see Ch.~11.5 of Ref.~\cite{Nrecipe} for details about this reformation) as a workaround to find
eigenvalues and eigenvectors using the function.

We compare the average time consumption to find all the eigenvectors of a
random $100\times100$ Hermitian matrix $A$ with a unit Frobenius norm for several different
internal precisions. To generate $A$, matrix elements $a_{ij}=a^*_{ji}$ are simply
randomly generated and then the normalization of the matrix is performed.
The eigenvectors found by a diagonalization are the column vectors of unitary matrix $U$ such
that $U^{-1}AU={\rm diag}$.

The probability for a random matrix to be degenerate is very small. Tested matrices were
in fact nondegenerate and ``\verb|eigen|'' worked fine except for the precision 256 [bits] for
which it stopped due to a low-accuracy error.
As for ``\verb|jacobi|'', we used the workaround as mentioned above so that the matrix actually
input into the function was a $200\times200$ real symmetric matrix. In this case, we
had to double the precision of the matrix in order to keep the off-diagonal elements
of $\widetilde{D}=U^{-1}AU$ sufficiently small ({\em i.e.},
$|\widetilde{D}_{ij, i\not = j}|\lesssim 2^{-({\rm prec}-10)}$) for the
specified precision ${\rm prec}$. This was not a theoretical consequence but an empirical
workaround for the use of ``\verb|jacobi|''. In fact, with precision 512 (1024) [bits], ``\verb|jacobi|''
constantly achieved $|\widetilde{D}_{ij, i\not = j}|\sim 1.0\times 10^{-78}\sim 1.0\times 2^{-256}$
($|\widetilde{D}_{ij, i\not = j}| \sim 1.0\times 10^{-155}\sim 1.0\times 2^{-512}$).
Thus, it was reasonable that doubled precision was handed over to ``\verb|jacobi|'' when it
was called.

As shown in Table\ \ref{tableHerm}, time consumption for ``\verb|diag_H|'' of ZKCM is on the
same order of magnitude as those for the functions ``\verb|eigen|'' and ``\verb|jacobi|'' of PARI,
as far as we tested for the precision between $256$ and $1280$ [bits].

In addition, we also compare the average time consumption to find all the eigenvectors
of a random $N\times N$ Hermitian matrix with a unit Frobenius norm for several different values of $N$ with
fixed precision $768$ [bits]. We doubled the precision for ``\verb|jacobi|'' due to the above-mentioned
reason. As shown in Table\ \ref{tableHerm2}, time consumption of ``\verb|diag_H|'' is on the
same order of magnitude as those for ``\verb|eigen|'' and ``\verb|jacobi|'' as far as we tested
for $25\le N \le 125$.
\begin{table}[pbt]
\caption{\label{tableHerm}Comparison of the average real time consumption to find all the
eigenvectors of a normalized random $100\times100$ Hermitian matrix. The average was taken
over ten different matrices. The standard deviation is shown in parentheses in small fonts.
``prec'' stands for the precision.
Environments: Fedora 15 OS on Intel Core i5 M460 CPU with 4GB RAM for upper entries ($\blacklozenge$) and
Fedora 16 OS on AMD FX-8120 CPU with 16GB RAM for lower entries ($\clubsuit$). Note: For precision 256 [bits],
function ``{\tt eigen}'' of PARI stopped with an error and output no result.
${}^*$ Precision was doubled for``{\tt jacobi}'' (see the text).}
\begin{center}
\begin{tabular}{|r|p{1.8cm}|p{1.8cm}|p{1.8cm}|}\hline
prec${}^*$&ZKCM, diag\_H [sec]& PARI, eigen [sec] & PARI, jacobi [sec]\\\hline
 256 & $\begin{array}{l}{\hspace{-2.15mm}{}^\blacklozenge}175\text{\scriptsize~(0.105)}\\{\hspace{-2.3mm}{}^\clubsuit}137\text{\scriptsize~(0.769)}\end{array}$ & $\begin{array}{l}{\rm N/A}\text{\scriptsize~(N/A)}\\{\rm N/A}\text{\scriptsize~(N/A)}\end{array}$ & $\begin{array}{l}103\text{\scriptsize~(0.324)}\\91.2\text{\scriptsize~(0.372)}\end{array}$\\\hline
 512 & $\begin{array}{l}259\text{\scriptsize~(0.160)}\\203\text{\scriptsize~(1.41)}\end{array}$ & $\begin{array}{l}171\text{\scriptsize~(1.27)}\\145\text{\scriptsize~(0.426)}\end{array}$ & $\begin{array}{l}265\text{\scriptsize~(0.974)}\\202\text{\scriptsize~(0.502)}\end{array}$\\\hline
 768 & $\begin{array}{l}413\text{\scriptsize~(0.378)}\\330\text{\scriptsize~(4.14)}\end{array}$ & $\begin{array}{l}237\text{\scriptsize~(1.24)}\\206\text{\scriptsize~(0.764)}\end{array}$ & $\begin{array}{l}477\text{\scriptsize~(2.10)}\\367\text{\scriptsize~(5.43)}\end{array}$\\\hline
1024 & $\begin{array}{l}632\text{\scriptsize~(0.358)}\\501\text{\scriptsize~(4.14)}\end{array}$ & $\begin{array}{l}378\text{\scriptsize~(1.58)}\\309\text{\scriptsize~(1.14)}\end{array}$ & $\begin{array}{l}726\text{\scriptsize~(2.24)}\\506\text{\scriptsize~(3.65)}\end{array}$\\\hline
1280 & $\begin{array}{l}903\text{\scriptsize~(0.315)}\\704\text{\scriptsize~(3.53)}\end{array}$ & $\begin{array}{l}503\text{\scriptsize~(1.21)}\\404\text{\scriptsize~(1.13)}\end{array}$ & $\begin{array}{l}1020\text{\scriptsize~(5.47)}\\734\text{\scriptsize~(7.98)}\end{array}$\\\hline
\end{tabular}
\end{center}
\end{table}
\begin{table}[pbt]
\caption{\label{tableHerm2}Comparison of the average real time consumption to find all the
eigenvectors of a normalized random $N\times N$ Hermitian matrix under the fixed precision
768 [bits] [precision was doubled for``{\tt jacobi}'' (see the text)]. The average was taken over
ten different matrices. The standard deviation is shown in parentheses in small fonts.
Environments: Fedora 15 OS on Intel Core i5 M460 CPU with 4GB RAM for upper entries ($\blacklozenge$) and
Fedora 16 OS on AMD FX-8120 CPU with 16GB RAM for lower entries ($\clubsuit$).}
\begin{center}
\begin{tabular}{|r|p{2.0cm}|p{2.0cm}|p{2.0cm}|}\hline
$N$ & ZKCM, diag\_H [sec]& PARI, eigen [sec] & PARI, jacobi [sec]\\\hline
 25 & $\begin{array}{l}{\hspace{-1.9mm}{}^\blacklozenge}3.41\text{\scriptsize~(0.0152)}\\{\hspace{-2.2mm}{}^\clubsuit}2.68\text{\scriptsize~(0.0130)}\end{array}$ & $\begin{array}{l}0.941\text{\scriptsize~(0.00406)}\\0.841\text{\scriptsize~(0.0399)}\end{array}$ & $\begin{array}{l}5.99\text{\scriptsize~(0.0699)}\\4.79\text{\scriptsize~(0.0219)}\end{array}$\\\hline
 50 & $\begin{array}{l}34.5\text{\scriptsize~(0.0624)}\\27.1\text{\scriptsize~(0.174)}\end{array}$ & $\begin{array}{l}14.5\text{\scriptsize~(0.0248)}\\12.8\text{\scriptsize~(0.0483)}\end{array}$ & $\begin{array}{l}50.9\text{\scriptsize~(0.364)}\\40.9\text{\scriptsize~(0.114)}\end{array}$\\\hline
 75 & $\begin{array}{l}146\text{\scriptsize~(0.238)}\\114\text{\scriptsize~(0.604)}\end{array}$ & $\begin{array}{l}74.3\text{\scriptsize~(0.636)}\\64.9\text{\scriptsize~(0.345)}\end{array}$ & $\begin{array}{l}182\text{\scriptsize~(0.977)}\\144\text{\scriptsize~(0.390)}\end{array}$\\\hline
100 & $\begin{array}{l}413\text{\scriptsize~(0.378)}\\330\text{\scriptsize~(4.14)}\end{array}$ & $\begin{array}{l}237\text{\scriptsize~(1.24)}\\206\text{\scriptsize~(0.764)}\end{array}$ & $\begin{array}{l}477\text{\scriptsize~(2.10)}\\367\text{\scriptsize~(5.43)}\end{array}$\\\hline
125 & $\begin{array}{l}961\text{\scriptsize~(1.10)}\\752\text{\scriptsize~(4.73)}\end{array}$ & $\begin{array}{l}596\text{\scriptsize~(1.72)}\\508\text{\scriptsize~(2.14)}\end{array}$ & $\begin{array}{l}1070\text{\scriptsize~(7.40)}\\754\text{\scriptsize~(4.62)}\end{array}$\\\hline
\end{tabular}
\end{center}
\end{table}

The evaluated functions are all designed to find eigenvectors in addition to eigenvalues. One may
think of the case only eigenvalues are needed, for which shorter computation time is expected. The
PARI library does not have a function for this purpose. The ZKCM library has the function
``\verb|eigenvalues_H|'' which is created by simply omitting the inverse iterations from ``\verb|diag_H|''.
The problem is that the precision of computed eigenvalues cannot be guaranteed without corresponding
eigenvectors. The achievable precision in the absence of inverse iterations is evaluated as below
together with the time consumption.

\paragraph{Numerical error in the absence of inverse iterations}
The performance of ``\verb|eigenvalues_H|'' is evaluated in Tables \ref{tableEI1} and \ref{tableEI2} in
comparison with ``\verb|diag_H|''. The real time consumption to find all the eigenvalues of a normalized
randomly-generated $N\times N$ Hermitian matrix and the numerical error in the computed eigenvalues are
used for this comparison. Here, the matrix was generated by a random unitary transformation of a
diagonal matrix $D$ with random real elements, where $D$ was normalized by its Frobenius norm in advance.
The numerical error in the computed spectrum $\{{e'}_i\}$ for ``\verb|eigenvalues_H|'' is quantified by
${\rm E}(\{{e'}_i\})=\sum_i|e_i - {e'}_i|$ where $e_i$ are the true eigenvalues. Here, \{${e'}_i\}$ and $\{e_i\}$
are sorted in the same order (we employed the descending order). Similarly, the numerical error in the computed
spectrum $\{{e''}_i\}$ for ``\verb|diag_H|'' is quantified by ${\rm E}(\{{e''}_i\})=\sum_i|e_i - {e''}_i|$.
For each cell of the tables, we performed the same simulation ten times with different random matrices and
took the average for the time consumption or the numerical error.
Note that this performance evaluation was separately performed with the evaluations shown in the previous tables.
The matrix size $N$ was fixed to $100$ and several values of precision were tried for Table \ref{tableEI1};
the precision was fixed to 768 [bits] and several values of $N$ were tried for Table \ref{tableEI2}.
\begin{table}[pbt]
\caption{\label{tableEI1}Comparison of the average real time consumption for ``{\tt eigenvalues\_H}''
and ``{\tt diag\_H}'' to find all the eigenvalues of a normalized random $100\times100$
Hermitian matrix. The average numerical error in the computed spectrum is also shown in the columns
``Error''. The average was taken over ten different matrices. In each cell, the standard deviation is
shown in parentheses in small fonts. ``prec'' stands for the precision.
Environments: Fedora 15 OS on Intel Core i5 M460 CPU with 4GB RAM for upper entries ($\blacklozenge$) and
Fedora 16 OS on AMD FX-8120 CPU with 16GB RAM for lower entries ($\clubsuit$).}
\begin{center}
\begin{tabular}{|p{0.55cm}|p{1.1cm}p{1.9cm}||p{1.0cm}p{2.1cm}|}\hline
~&{\tt eigenvalues\_H}&~&{\tt diag\_H}&~
\end{tabular}~\\
\vspace{-1mm}
\begin{tabular}{|p{0.55cm}|p{1.1cm}|p{1.9cm}||p{1.0cm}|p{2.1cm}|}\hline
prec&Time [sec]& Error & Time [sec] & Error\\\hline
 256 & $\begin{array}{l}{\hspace{-2.15mm}{}^\blacklozenge}9.24\\\text{\scriptsize~(0.250)}\\{\hspace{-2.3mm}{}^\clubsuit}7.26\\\text{\scriptsize~(0.149)}\end{array}$ & $\begin{array}{l}1.70\times10^{-32}\\\text{\scriptsize~($2.98\times10^{-32}$)}\\1.19\times10^{-32}\\\text{\scriptsize~($2.00\times10^{-32}$)}\end{array}$ & $\begin{array}{l}173\\\text{\scriptsize~(1.05)}\\136\\\text{\scriptsize~(1.13)}\end{array}$ & $\begin{array}{l}7.24\times10^{-76}\\\text{\scriptsize~($8.69\times10^{-76}$)}\\3.83\times10^{-76}\\\text{\scriptsize~($2.01\times10^{-76}$)}\end{array}$\\\hline
 512 & $\begin{array}{l}13.2\\\text{\scriptsize~(0.194)}\\10.4\\\text{\scriptsize~(0.142)}\end{array}$ & $\begin{array}{l}1.32\times10^{-49}\\\text{\scriptsize~($2.99\times10^{-49}$)}\\3.63\times10^{-49}\\\text{\scriptsize~($6.70\times10^{-49}$)}\end{array}$ & $\begin{array}{l}259\\\text{\scriptsize~(1.60)}\\202\\\text{\scriptsize~(1.88)}\end{array}$ & $\begin{array}{l}3.86\times10^{-153}\\\text{\scriptsize~($1.26\times10^{-153}$)}\\3.96\times10^{-153}\\\text{\scriptsize~($2.00\times10^{-153}$)}\end{array}$\\\hline
 768 & $\begin{array}{l}19.1\\\text{\scriptsize~(0.368)}\\15.2\\\text{\scriptsize~(0.272)}\end{array}$ & $\begin{array}{l}9.35\times10^{-57}\\\text{\scriptsize~($8.62\times10^{-57}$)}\\3.71\times10^{-56}\\\text{\scriptsize~($8.16\times10^{-56}$)}\end{array}$ & $\begin{array}{l}413\\\text{\scriptsize~(2.61)}\\327\\\text{\scriptsize~(2.82)}\end{array}$ & $\begin{array}{l}3.01\times10^{-230}\\\text{\scriptsize~($1.12\times10^{-230}$)}\\4.88\times10^{-230}\\\text{\scriptsize~($5.43\times10^{-230}$)}\end{array}$\\\hline
1024 & $\begin{array}{l}27.6\\\text{\scriptsize~(0.454)}\\22.1\\\text{\scriptsize~(0.346)}\end{array}$ & $\begin{array}{l}2.08\times10^{-56}\\\text{\scriptsize~($3.71\times10^{-56}$)}\\4.00\times10^{-56}\\\text{\scriptsize~($7.49\times10^{-56}$)}\end{array}$ & $\begin{array}{l}637\\\text{\scriptsize~(3.27)}\\498\\\text{\scriptsize~(2.99)}\end{array}$ & $\begin{array}{l}4.03\times10^{-307}\\\text{\scriptsize~($3.58\times10^{-307}$)}\\3.55\times10^{-307}\\\text{\scriptsize~($3.01\times10^{-307}$)}\end{array}$\\\hline
1280 & $\begin{array}{l}35.9\\\text{\scriptsize~(0.871)}\\28.4\\\text{\scriptsize~(0.539)}\end{array}$ & $\begin{array}{l}6.00\times10^{-56}\\\text{\scriptsize~($9.69\times10^{-56}$)}\\9.84\times10^{-54}\\\text{\scriptsize~($3.10\times10^{-53}$)}\end{array}$ & $\begin{array}{l}904\\\text{\scriptsize~(4.00)}\\706\\\text{\scriptsize~(4.64)}\end{array}$ & $\begin{array}{l}6.65\times10^{-384}\\\text{\scriptsize~($1.18\times10^{-383}$)}\\2.11\times10^{-384}\\\text{\scriptsize~($6.12\times10^{-385}$)}\end{array}$\\\hline
\end{tabular}
\end{center}
\end{table}
\begin{table}[pbt]
\caption{\label{tableEI2}Comparison of ``{\tt eigenvalues\_H}'' and ``{\tt diag\_H}'' in real time consumption
to find all the eigenvalues of a normalized random $N\times N$ Hermitian matrix, for several different matrix
sizes $N$ for the fixed precision $768$ [bits]. The average over ten trials is shown. The columns ``Error''
show the average numerical error in the computed spectrum. The standard deviation is shown in parentheses for
each value. Environments: Fedora 15 OS on Intel Core i5 M460 CPU with 4GB RAM for upper entries
($\blacklozenge$) and Fedora 16 OS on AMD FX-8120 CPU with 16GB RAM for lower entries ($\clubsuit$).}
\begin{center}
\begin{tabular}{|p{0.55cm}|p{1.1cm}p{1.9cm}||p{1.0cm}p{2.1cm}|}\hline
~&{\tt eigenvalues\_H}&~&{\tt diag\_H}&~
\end{tabular}~\\
\vspace{-1mm}
\begin{tabular}{|p{0.55cm}|p{1.1cm}|p{1.9cm}||p{1.0cm}|p{2.1cm}|}\hline
$N$&Time [sec]& Error & Time [sec] & Error\\\hline
 25 & $\begin{array}{l}{\hspace{-2.15mm}{}^\blacklozenge}0.438\\\text{\scriptsize~(0.0179)}\\{\hspace{-2.3mm}{}^\clubsuit}0.343\\\text{\scriptsize~(0.0153)}\end{array}$ & $\begin{array}{l}4.46\times10^{-57}\\\text{\scriptsize~($8.47\times10^{-57}$)}\\3.42\times10^{-57}\\\text{\scriptsize~($8.76\times10^{-57}$)}\end{array}$ & $\begin{array}{l}3.34\\\text{\scriptsize~(0.0253)}\\2.69\\\text{\scriptsize~(0.0302)}\end{array}$ & $\begin{array}{l}9.10\times10^{-231}\\\text{\scriptsize~($3.84\times10^{-231}$)}\\6.35\times10^{-231}\\\text{\scriptsize~($1.72\times10^{-231}$)}\end{array}$\\\hline
 50 & $\begin{array}{l}2.75\\\text{\scriptsize~(0.0704)}\\2.19\\\text{\scriptsize~(0.0824)}\end{array}$ & $\begin{array}{l}1.67\times10^{-56}\\\text{\scriptsize~($3.14\times10^{-56}$)}\\9.67\times10^{-57}\\\text{\scriptsize~($1.49\times10^{-56}$)}\end{array}$ & $\begin{array}{l}34.0\\\text{\scriptsize~(0.220)}\\27.2\\\text{\scriptsize~(0.190)}\end{array}$ & $\begin{array}{l}2.28\times10^{-230}\\\text{\scriptsize~($2.51\times10^{-230}$)}\\2.08\times10^{-230}\\\text{\scriptsize~($1.02\times10^{-230}$)}\end{array}$\\\hline
 75 & $\begin{array}{l}8.44\\\text{\scriptsize~(0.149)}\\6.73\\\text{\scriptsize~(0.163)}\end{array}$ & $\begin{array}{l}7.30\times10^{-56}\\\text{\scriptsize~($1.58\times10^{-55}$)}\\2.64\times10^{-56}\\\text{\scriptsize~($3.56\times10^{-56}$)}\end{array}$ & $\begin{array}{l}144\\\text{\scriptsize~(0.762)}\\114\\\text{\scriptsize~(0.692)}\end{array}$ & $\begin{array}{l}3.16\times10^{-230}\\\text{\scriptsize~($3.58\times10^{-230}$)}\\2.40\times10^{-230}\\\text{\scriptsize~($1.03\times10^{-230}$)}\end{array}$\\\hline
 100 & $\begin{array}{l}19.1\\\text{\scriptsize~(0.368)}\\15.2\\\text{\scriptsize~(0.272)}\end{array}$ & $\begin{array}{l}9.35\times10^{-57}\\\text{\scriptsize~($8.62\times10^{-57}$)}\\3.71\times10^{-56}\\\text{\scriptsize~($8.16\times10^{-56}$)}\end{array}$ & $\begin{array}{l}413\\\text{\scriptsize~(2.61)}\\327\\\text{\scriptsize~(2.82)}\end{array}$ & $\begin{array}{l}3.01\times10^{-230}\\\text{\scriptsize~($1.12\times10^{-230}$)}\\4.88\times10^{-230}\\\text{\scriptsize~($5.43\times10^{-230}$)}\end{array}$\\\hline
125 & $\begin{array}{l}36.4\\\text{\scriptsize~(0.794)}\\28.4\\\text{\scriptsize~(0.478)}\end{array}$ & $\begin{array}{l}1.22\times10^{-55}\\\text{\scriptsize~($2.28\times10^{-55}$)}\\1.52\times10^{-56}\\\text{\scriptsize~($2.08\times10^{-56}$)}\end{array}$ & $\begin{array}{l}950\\\text{\scriptsize~(5.58)}\\753\\\text{\scriptsize~(5.25)}\end{array}$ & $\begin{array}{l}3.76\times10^{-230}\\\text{\scriptsize~($1.75\times10^{-230}$)}\\3.15\times10^{-230}\\\text{\scriptsize~($8.01\times10^{-231}$)}\end{array}$\\\hline
\end{tabular}
\end{center}
\end{table}

It has turned out that ``\verb|eigenvalues_H|'' is faster than ``\verb|diag_H|'' by one or two orders of
magnitude to compute eigenvalues. The only difference in the internal structures of these functions is
the inverse iterations to find eigenvectors and at the same time enhance the accuracy of eigenvalues.
Thus the result shows most of time consumption in ``\verb|diag_H|'' is spent for inverse iterations.
A drawback to use ``\verb|eigenvalues_H|'' is the nonnegligible error in the computed eigenvalues.
For instance, we have $<{\rm E}(\{{e'}_i\})>\sim 10^{-49}$ when the precision is $512$ [bits], {\em i.e.},
when the machine epsilon is $\sim 10^{-154}$. Indeed, the error is much smaller than the machine epsilon
for double precision ($\sim 10^{-16}$), but is not in the acceptable range for multiprecision computation.

So far, the ZKCM library has been introduced and evaluated on its performance.
An extension library for the study of quantum computation will be introduced
in the next section.

\section{ZKCM\_QC library}\label{secZKCMQC}
The ZKCM\_QC library is an extension of the ZKCM library.
It has several classes to handle tensor data useful for
the (time-dependent) MPS simulation of a quantum circuit.
Among the classes, the ``\verb|mps|'' class and the ``\verb|tensor2|'' class
will be used by user-side programs. The former class
conceals the complicated MPS simulation process and enables 
writing programs in a simple manner. The latter class is
used to represent two-dimensional tensors which are often 
simply regarded as matrices. A quantum state during an MPS
simulation can be obtained as a (reduced) density matrix in the
type of ``\verb|tensor2|''. For convenience, there are functions to
convert a matrix in the type of ``\verb|tensor2|'' to the type of
``\verb|zkcm_matrix|'' and {\em vice versa}.
More details of the classes are explained in the document placed
in the ``doc'' directory of the ZKCM\_QC package.

It might be curious to condensed-matter physicists why multiprecision computation
is needed in an MPS simulation. Indeed, truncations of Schmidt coefficients (and corresponding
Schmidt vectors) have been studied as a possible source of errors \cite{VDMB09} while precision
of basic floating-point operations has not been of main concern in physics simulations using
the MPS data structure. Precision shortage cannot be a main source of errors in light of
truncation errors. However, it has been uncommon\footnote{Here, we are considering the standard quantum
circuit model for quantum computing. It is not the case in a Hamiltonian-based adiabatic-evolution model
\cite{Banuls06}.} to use a truncation of nonzero Schmidt coefficients in time-dependent MPS simulations
of quantum computing \cite{V03,KW04,S06}. Under this condition, precision shortage becomes the only
crucial source of errors and it is hence of our main concern how much precision is required for an
accurate simulation of quantum computing. We will discuss more on this issue in Sec.~\ref{secDJ}. In
particular, we will show an example of simulating a quantum algorithm for which a truncation of at
most a single nonzero Schmidt coefficient for each step results in a significant error; in
addition, a precision slightly beyond the double precision is necessary for this example.

As for installation of the ZKCM\_QC library, the standard process
``\verb|./configure| $\rightarrow$ \verb|make| $\rightarrow$ \verb|sudo make install|''
should work; if not, the document should be consulted.

We briefly overview the theory of the MPS simulation before
introducing a program example.
\subsection{Brief overview of the theory of time-dependent MPS simulation}\label{sectheory}
Consider an $n$-qubit pure quantum state
\[
|\Psi\rangle=\sum_{i_0\cdots i_{n-1}=0\cdots0}^{1\cdots1}
c_{i_0\cdots i_{n-1}}|i_0\cdots i_{n-1}\rangle
\]
with $\sum_{i_0\cdots i_{n-1}}|c_{i_0\cdots i_{n-1}}|^2=1$.
If we keep this state as data as it is, updating the data for each
time of unitary time evolution spends $O(2^{2n})$ floating-point
operations. To avoid such an exhaustive calculation, in the
matrix-product-state method, the data is stored as a kind of compressed
data. The state is represented in the form
\begin{equation}\label{eq1}
\begin{split}
 |\Psi\rangle&=\sum_{i_0=0}^1\cdots\sum_{i_{n-1}=0}^1\biggl[
\sum_{v_0=0}^{m_0-1}\sum_{v_1=0}^{m_1-1}\cdots\sum_{v_{n-2}=0}^{m_{n-2}-1}\\
&~~Q_0(i_0,v_0)V_0(v_0)Q_1(i_1,v_0,v_1)V_1(v_1)\cdots \\&~~\cdots 
Q_{s}(i_{s},v_{s-1},v_s)V_s(v_s)\cdots\\&~~\cdots V_{n-2}(v_{n-2})Q_{n-1}(i_{n-1},v_{n-2})\biggr]
|i_0\cdots i_{n-1}\rangle,
\end{split}\end{equation}
where we use tensors $\{Q_s\}_{s=0}^{n-1}$ with parameters $i_s, v_{s-1}, v_s$
(parameters $v_{-1}$ and $v_{n-1}$ exceptionally do not exist) and $\{V_s\}_{s=0}^{n-2}$
with parameter $v_s$; $m_s$ is a suitable number of values assigned to $v_s$
with which the state is represented precisely or well approximated.
This form is one of the forms of matrix product states (MPSs). The
data are kept in the tensors. We can see that neighboring
tensors are correlated to each other; the data compression is owing
to this structure.

Let us explain a little more details:
$Q_s(i_s,v_{s-1},v_{s})$ is a tensor with
$2\times m_{s-1} \times m_{s}$ elements; $V_s(v_s)$ is a tensor
in which the Schmidt coefficients for the splitting between the $s$th
site and the $(s+1)$th site ({\em i.e.}, the positive square roots of nonzero
eigenvalues of the reduced density operator of qubits $0,\ldots,s$) are
stored. This implies that, by using $V_s$ and eigenvectors
$|\Phi_{v_s}^{0\ldots s}\rangle$ ($|\Phi_{v_s}^{s+1\ldots n-1}\rangle$)
of the reduced density operator $\rho^{0\ldots s}$ ($\rho^{s+1\ldots n-1}$)
of qubits $0,\ldots,s$ ($s+1,\ldots,n-1$), the state can also be written
in the form of Schmidt decomposition
\begin{equation}
 |\Psi\rangle = \sum_{v_s=0}^{m_s-1}V_s(v_s)
|\Phi_{v_s}^{0\ldots s}\rangle |\Phi_{v_s}^{s+1\ldots n-1}\rangle.
\end{equation}
In an MPS simulation, very small coefficients and corresponding
eigenvectors can be truncated out unlike a usual Schmidt decomposition.
It may happen that all the nonzero coefficients are nonnegligible.
This is actually the case in practice when we simulate quantum computing
as we will discuss in Sec.~\ref{secDJ}. We can still enjoy a considerable
data-size reduction since vanishing coefficients are truncated out.

Besides the truncation, a significant advantage to use the MPS form is
that we have only to handle a small number of tensors when we simulate a time
evolution under each quantum gate. For example, when we apply a unitary
operation $\in {\rm U}(4)$ acting on, say, qubits $s$ and $s+1$,
we have only to update the tensors $Q_s(i_s,v_{s-1},v_s)$, $V_s(v_s)$,
and $Q_{s+1}(i_{s+1},v_s,v_{s+1})$.
For the details of how tensors are updated, see Refs.\ \cite{V03,S06}.
The simulation of a single quantum gate $\in {\rm U}(4)$ spends
$O(m_{\rm max}^3)$ floating-point operations where $m_{\rm max}$
is the largest value of $m_s$ among the sites $s$.
(Usually, unitary operations $\in {\rm U}(2)$ and those $\in {\rm U}(4)$
are regarded as elementary quantum gates.)
A quantum circuit constructed by using at most $g$
single-qubit and/or two-qubit quantum gates can be simulated within the
cost of $O(g n m_{\rm max,max}^3)$ floating-point operations,
where $n$ is the number of wires and $m_{\rm max,max}$ is the largest
value of $m_{\rm max}$ over all time steps. 

The computational complexity may be slightly different for each software
using MPS. In the ZKCM\_QC library, we have functions to apply quantum gates
$\in {\rm U}(8)$ to three chosen qubits. Internally, three-qubit gates
are handled as elementary gates. This makes the complexity a little
larger. A simulation using the library spends $O(g n m_{\rm max,max}^4)$
floating-point operations, where $g$ is the number of single-qubit, two-qubit,
and/or three-qubit gates used for constructing a quantum circuit. As the
process to update the tensors to simulate a three-qubit gate has not been
written in detail in the literature as far as the author knows, it is explained
in \ref{appendix}.

The MPS simulation process, which is in fact often complicated, can be
concealed by the use of ZKCM\_QC. One may write a program for quantum circuit
simulation in an intuitive manner. Here is a very simple example. In the
followings, we use version 0.1.0 of ZKCM\_QC for our programs.

\subsection{Program example}\label{secprogexample}
As a simple example, we simulate the time evolution
\[\begin{split}
&|000\rangle\overset{H}{\mapsto} \frac{1}{\sqrt{2}}(|000\rangle+|100\rangle)\\
&\overset{\rm CNOT}{\mapsto} \frac{1}{\sqrt{2}}(|000\rangle+|101\rangle)
\end{split}\]
where Hadamard operation $H=\frac{1}{\sqrt{2}}\begin{pmatrix}1&1\\1&-1\end{pmatrix}$
acts on qubit 0 and ${\rm CNOT}=|00\rangle\langle00|+|01\rangle\langle01|
+|10\rangle\langle11|+|11\rangle\langle10|$ acts on qubits 0 and 2.
Then we see the reduced density matrix of qubits 0 and 2.

The program example is shown in Listing~\ref{codempssimple}. (A slightly extended
sample code is placed in the ``samples'' directory of the ZKCM\_QC package.) It
utilizes several matrices declared in the namespace ``\verb|tensor2tools|'' (see the
document placed in the ``doc'' directory for the details of this namespace).
\begin{lstlisting}[label=codempssimple,caption=qc\_simple\_example.cpp]
#include "zkcm_qc.hpp"

int main (int argc, char *argv[])
{
  //Use the 256-bit precision.
  zkcm_set_default_prec(256);
  //Num. of digits for each output is set to 8.
  zkcm_set_output_nd(8);

  //First, we make an MPS representing |000>.
  mps M(3);
  std::cout << "The initial state is "
            << std::endl;
  //Print the reduced density operator of the
  //block of qubits from 0 to 2, namely, 0,1,2,
  //using the binary number representation for
  //basis vectors.
  std::cout << M.RDO_block(0, 2).str_dirac_b()
            << std::endl;

  std::cout << "Now we apply H to the 0th qubit."
            << std::endl;
  M.applyU(tensor2tools::Hadamard, 0);

  std::cout<< "Now we apply CNOT to the \
qubits 0 and 2."
           << std::endl;
  M.applyU(tensor2tools::CNOT, 0, 2);

  //The array is used to specify qubits to compute
  //a reduced density matrix. It should be
  //terminated by the constant mps::TA.
  int array[] = {0, 2, mps::TA};
  std::cout << "At this point, the reduced \
density matrix of the qubits 0 and 2 is "
            << std::endl
            << M.RDO(array).str_dirac_b()
            << std::endl;
  zkcm_quit();
  return 0;
}
\end{lstlisting}
The program is compiled and executed typically in the following way.
\begin{Verbatim}
[user@localhost foo]$ c++ -o qc_simple_example \
qc_simple_example.cpp -lzkcm_qc -lzkcm -lmpfr -lgmp \
-lgmpxx
[user@localhost foo]$ ./qc_simple_example
The initial state is
1.0000000e+00|000><000|
Now we apply H to the 0th qubit.
Now we apply CNOT to the qubits 0 and 2.
At this point, the reduced density matrix of \
the qubits 0 and 2 is
5.0000000e-01|00><00|+5.0000000e-01|00><11|\
+5.0000000e-01|11><00|+5.0000000e-01|11><11|
\end{Verbatim}

Besides this example, the package of ZKCM\_QC has sample programs for
simulating Grover's quantum search \cite{Gr96} and simulating a simple projective
measurement. In addition, we will see an example to simulate the quantum
search under the simplest setting in \ref{simplestGrover},
which is written as a part of the explanation of how to handle a three-qubit gate
in an MPS simulation.

\subsection{Source of numerical errors in an MPS simulation of quantum computing}\label{secDJ}
It is a standard strategy in the MPS simulation and related methods in computational
condensed matter physics to impose a certain threshold $m_{\rm trunc}$ to the number
of Schmidt coefficients; only larger $m_{\rm trunc}$ Schmidt coefficients (and corresponding
Schmidt vectors) are employed and the remaining are discarded at each time when tensors are
updated \cite{WH92,WH93}. It has been tacitly assumed that truncations are the main source
of numerical errors and a numerical error due to precision shortage has not gathered attention.
Venzl {\em et al.} \cite{VDMB09} pointed out that hardness of a time-dependent MPS simulation depends on
the distribution of Schmidt coefficients. They reported certain parameter choices in
the tilted Bose-Hubbard model, for which the distribution has a rather long tail for larger
Schmidt coefficients, {\em i.e.}, negligibly small ones are not dominant. In such a case,
$m_{\rm trunc}$ should be rather close to the maximum Schmidt rank (over all splittings between
consecutive sites and over all time steps) in order to keep the truncation error small. In contrast,
a usual time-dependent MPS simulation for other nearest-neighbour coupling models studied so far uses
a relatively small value for $m_{\rm trunc}$, typically fifty to one hundred \cite{Gob05}. With this much
value of $m_{\rm trunc}$, a time-dependent MPS simulation usually does not exhibit notable accumulation of
numerical errors unless more than several hundred time steps of evolution is tried \cite{Gob05}.
A significant accumulation of truncation errors were reported for a larger number of time steps
\cite{Gob05,JJGR06}.

As mentioned before, it has been uncommon to use truncations in (time-dependent) MPS simulations of quantum
computing \cite{V03,KW04,S06}. This is reasonable because quantum algorithms involves many
$H$ and ${\rm CNOT}$ operations (see Sec.~\ref{secprogexample} for their definitions). This makes a
quantum state evolving under a quantum circuit quite often a nearly equally biased superposition of
several indices. By tracing-out some subsystem under this condition, we have a reduced density matrix
whose nonzero eigenvalues are all nonnegligible. Thus it is not possible to truncate out even a single
nonzero Schmidt coefficient, irrespective of the number of time steps.
In addition, we have recently shown \cite{S12} that an accumulation of errors due to precision shortage
is significant even without truncation of nonzero Schmidt coefficients as far as we have seen in an
example to iterate the quantum search routine \cite{Gr96}. Thus high-precision computation has
a practical advantage in an MPS simulation for a quantum circuit model with a large circuit depth.

Here, we show a typical example where truncating out at most a single nonzero Schmidt coefficient
for each time causes a significant error. In this example, slightly more than double precision is
required for a stable simulation although the depth of the quantum circuit is not very large.

We consider the Deutsch-Jozsa algorithm \cite{DJ92}. In a brief explanation, there is a promise
that function $f:\{0,1\}^l\rightarrow\{0,1\}$ is either balanced ({\em i.e.},
$\#\{{\bf x}|f({\bf x})=0\}=\#\{{\bf x}|f({\bf x})=1\}$) or constant ({\em i.e.}, $f({\bf x})$ is
same for all ${\bf x}$) where ${\bf x}\in\{0_0\cdots0_{l-1},\ldots,1_0\cdots1_{l-1}\}$. The task is
to decide whether a given function $f$ is balanced or constant. This takes $1+2^{l-1}$ queries for
the worst case in classical computation while it takes only a single query in quantum computation
with the Deutsch-Jozsa algorithm.
The algorithm is described as follows: (i) Apply $H^{\otimes l}V_fH^{\otimes l}$ to $l$ qubits
initially in the state $|0_0\cdots0_{l-1}\rangle$ where $V_f$ is an operation to put the factor
$(-1)^{f({\bf x})}$ to each $|{\bf x}\rangle$; (ii) Measure the $l$ qubits in the computational
basis. When $f$ is balanced, the probability of finding the qubits simultaneously in $0$'s in
this measurement vanishes; when $f$ is constant, it is exactly unity. For the details of the
theoretical analysis of the algorithm, see, {\em e.g.}, Sec.~3.1.2 of Ref.~\cite{Gruska}.

\begin{figure}
\begin{center}
\scalebox{0.48}{\includegraphics{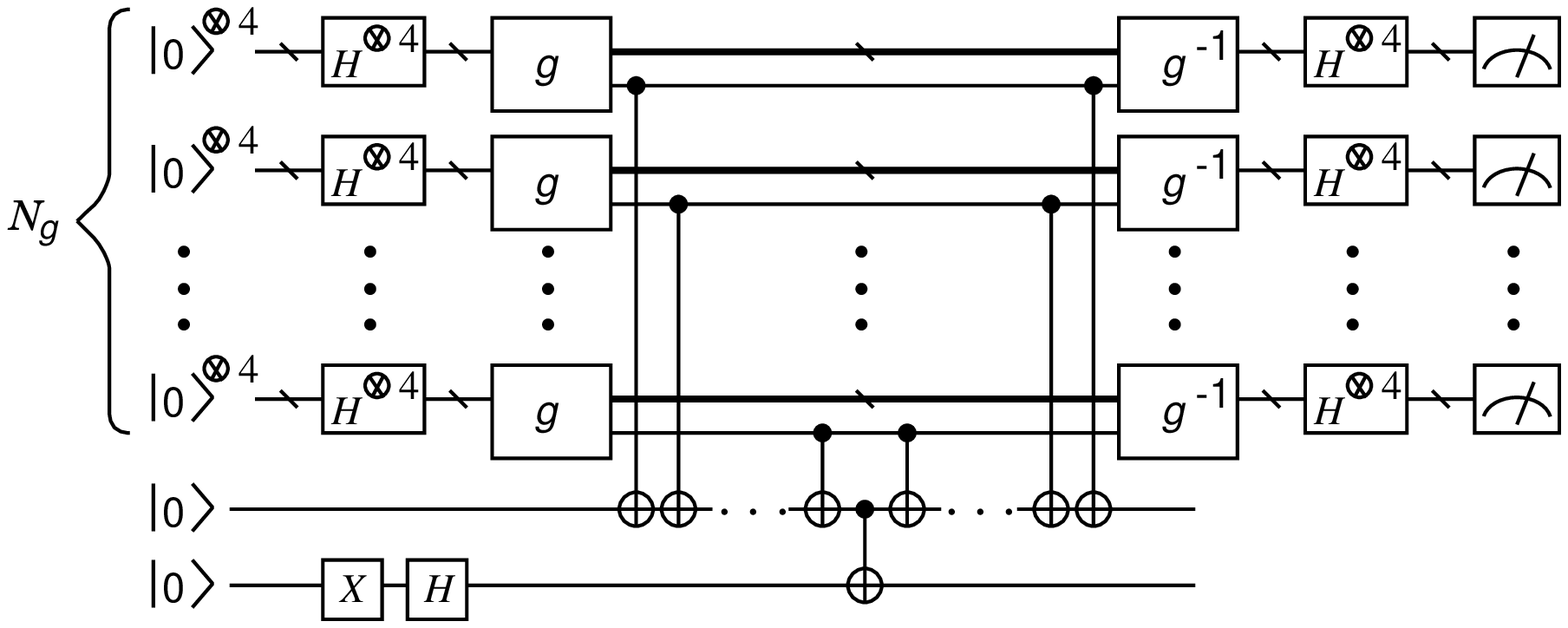}}\\(a)\\~\\
\scalebox{0.48}{\includegraphics{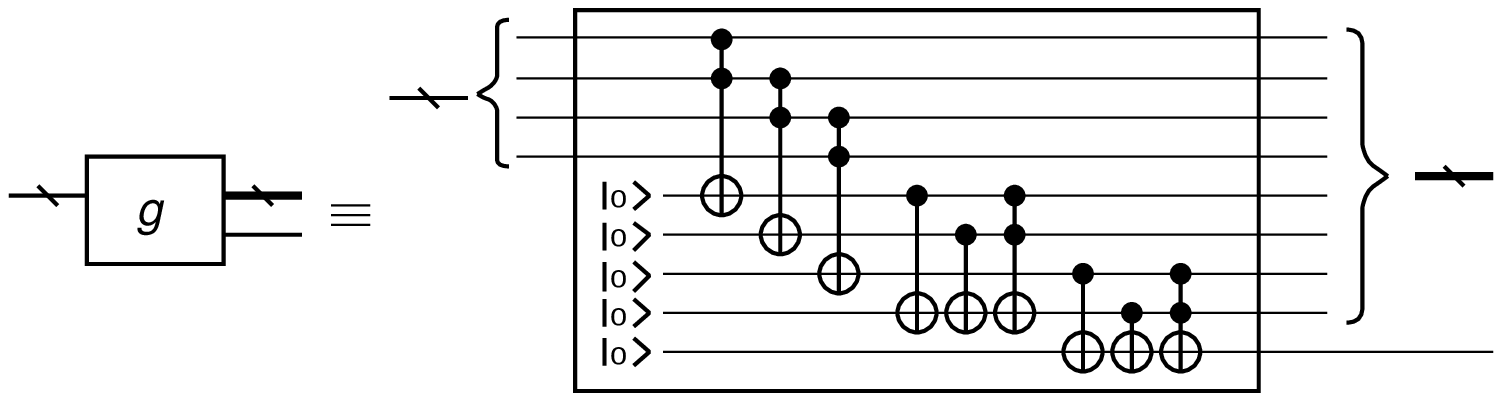}}\\(b)
\caption{\label{figDJcirc}(a) Quantum circuit of the Deutsch-Jozsa algorithm for
the specified function (see the text). (b) Internal structure of gate $g$.
Gate $g^{-1}$ is the reverse of $g$.}
\end{center}
\end{figure}

Let us consider a particular function $f({\bf y}^{}_0\cdots{\bf y}^{}_{N_g-1})
=\bigoplus_{i=0}^{N_g-1}g({\bf y}_i)$ with $g(x_0x_1x_2x_3)=(x_0\wedge x_1)\vee(x_1\wedge x_2)
\vee(x_2\wedge x_3)$ where ${\bf y}^{}_i\in\{0,1\}^4$ and $x_j\in\{0,1\}$; $N_g$ is a
positive integer (here, symbol $\bigoplus$ stands for the exclusive OR operation).
Figure \ref{figDJcirc} shows the quantum circuit of the algorithm for this function.
This function is balanced for any value of $N_g \ge 1$.
In the figure, a single $\bullet$ and the connected $\oplus$ represent the CNOT
operation with the control qubit specified by $\bullet$ and the target qubit specified by $\oplus$.
It flips the target qubit if the control qubit is in $|1\rangle$.
Similarly, two $\bullet$'s and $\oplus$ connected to each other represent the CCNOT
operation. It flips the target qubit if the two control qubits are in $|11\rangle$.
(See also \ref{appendix} for the implementation of three-qubit operations in our library.)
By the structure of the circuit, each of the $N_g$ measurements should report
${\rm Prob}(0000)=0$ if there is no numerical error. (This is easily proved:
assuming different values of ${\rm Prob}(0000)$ for two different bundles of qubits
contradicts to the fact that the bundles are equivalent to each other by the circuit
structure.)

It is straight-forward to write a program code for the quantum circuit using the ZKCM\_QC library.
We begin with header file descriptions:
\begin{Verbatim}
#include <zkcm_qc.hpp>
#include <sys/time.h>
\end{Verbatim}
The second header file is needed by the following function to obtain the current time in seconds:
\begin{Verbatim}
double current_time_in_sec ()
{
  timeval T;
  ::gettimeofday(&T, NULL);
  return ((double)(T.tv_sec)
          + 1.0e-6 * (double)(T.tv_usec));
}
\end{Verbatim}
We use this function when we record the data of time consumption for Table \ref{tableDJprec}.
We may omit it otherwise. The main function begins with
\begin{Verbatim}
int main (int argc, char *argv[])
{
  zkcm_set_default_prec (prec);
\end{Verbatim}
with some precision ``\verb|prec|'' and ends with
\begin{Verbatim}
  zkcm_quit();
  return 0;
}
\end{Verbatim}
We describe the quantum circuit step by step in the middle part of the main function.
(We call ``\verb|current_time_in_sec()|'' before and after this part to calculate the time
consumption.)
First of all, an object of the MPS data structure for $n=9N_g+2$ qubits is created by
\begin{Verbatim}
mps M(n);
M.set_m_trunc(mtrunc);
\end{Verbatim}
where we also set the value of $m_{\rm trunc}$ by our option as in the second line.
Then we write each gate operation one by one. For example, a Pauli $X$ gate (or a bit flip)
acting on the $i$th qubit is written as 
\begin{Verbatim}
M.applyU(tensor2tools::PauliX, i);
\end{Verbatim}
Similarly, an Hadamard gate operation is written in the same manner with \verb|tensor2tools::Hadamard|.
A CNOT gate with control qubit $c$ and target qubit $t$ is written as
\begin{Verbatim}
M.applyU(tensor2tools::CNOT, c, t);
\end{Verbatim}
A CCNOT operation with control qubits $a$, $b$, and target qubit $t$ is written as
\begin{Verbatim}
M.CCNOT(a, b, t);
\end{Verbatim}
After writing instructions corresponding to individual quantum gate operations, we need some code
lines to obtain the probability of finding zeros in the output. Although we have a function to
perform a projective measurement acting on a single qubit, this is not convenient. We instead
compute the $(0,0)$ element of the reduced density matrix of qubits of our interest. To obtain
${\rm Prob}(0_00_10_20_3)$, for example, we write
\begin{Verbatim}
std::cout << "Prob(0_0 0_1 0_2 0_3)="
          << M.RDO_block(0, 3).get(0, 0)
          << std::endl;
\end{Verbatim}
In this way, one can easily write a program code for the quantum circuit.

In the present context, it is also demanded to see the intrinsic data of the MPS step by step.
To obtain the number $m_s$ of surviving Schmidt coefficients $V_s(v_s)$ for the splitting between
sites $s$ and $s+1$, one may write, {\rm e.g.},
\begin{Verbatim}
int num = M.get_m(s);
\end{Verbatim}
for some integer $0\le s \le n - 2$. In addition, each Schmidt coefficient is accessible by
``\verb|M.V(s)(i)|'' which is a reference to the $i$th element of $V_s$. It is also convenient
to use ``\verb|M.get_m_max()|'' ($m_{\rm max}={\rm max}_s m_s$ for the current time step) and
``\verb|M.get_m_maxmax()|'' ($m_{\rm max,max}={\rm max}_t {\rm max}_s m_s$ where $t$ stands for time)
so as to find the time step where the Schmidt rank reaches the maximum. It should be noted that sometimes
the maximum Schmidt rank appears during the internal process hidden behind each gate operation. This
is because each gate operation for nonconsecutive qubits needs to internally move them to consecutive places
before operation and move them back to their original places after operation owing to the one-dimensional
data structure of MPS. This is done by swapping consecutive qubits step by step. For tracing the internal
time evolution of $m_s$'s, one needs to mimic this process by using the operation ``\verb|M.SWAP(a,b);|''
(this swaps specified qubits $a$ and $b$) manually.

\begin{figure}[pbt]
\begin{center}
\resizebox{0.5\textwidth}{!}{\includegraphics{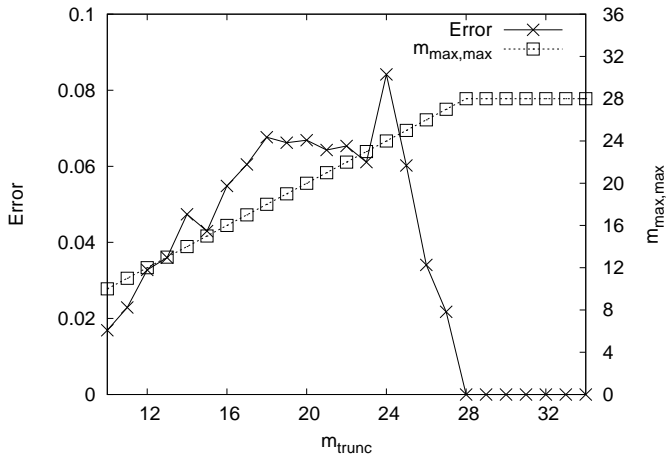}}
\caption{\label{figDJrank}Error in the computed value of ${\rm Prob}(0_00_10_20_3)$
and $m_{\rm max,max}$ as functions of $m_{\rm trunc}$. Precision was fixed to 256 bits.}
\end{center}
\end{figure}

We have explained how the program code is written to simulate the quantum circuit of Fig.~\ref{figDJcirc}.
Now we set $N_g$ to $7$; we have $65$ qubits in total in the circuit. It is certainly intractable
to handle the circuit by the brute-force method because the dimension of the Hilbert space is
$2^{65}\simeq 3.69\times10^{19}$. It was numerically found, however, that the largest possible Schmidt rank
of the MPS during evolution in the circuit is only $28$ as shown in Fig.\ \ref{figDJrank}.
Because of this reason, the MPS simulation of the algorithm took only approximately seven minutes (see
Table \ref{tableDJprec}) when the precision was set to $256$ [bits] and no truncation of nonzero Schmidt
coefficients was employed. Here, we used a computing server with the Red Hat Enterprise Linux 6 64-bit OS,
Intel Xeon E7-8837 2.67GHz (2.80GHZ maximum) CPU, and 315GB RAM, instead of the common PCs that we normally
used.\footnote{This was because a PC with 4GB RAM became unstable due to memory shortage for this
simulation for precision $\ge 1024$ bits. Each run of the simulation in the computing server consumed at
most approximately 8GB memory space for precision 1024 bits and 1280 bits.}

\begin{figure}[pbt]
\begin{center}
\resizebox{0.5\textwidth}{!}{\includegraphics{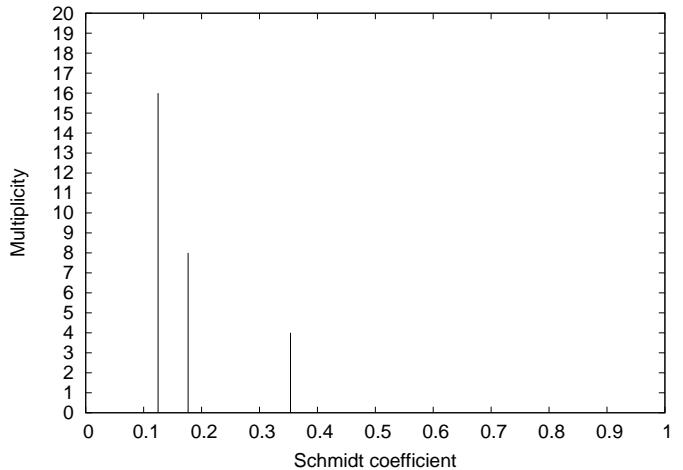}}
\caption{\label{figDist}Distribution of nonzero Schmidt coefficients at the point where
the second CNOT gate was being processed among the $2N_g+1$ CNOT gates ($N_g$ was set to 7) in the
middle part of Fig.~\ref{figDJcirc}(a). Note that we are employing the definition of
Schmidt coefficients with square roots in this paper. Thus the sum of squared values
of the Schmidt coefficients equals to one.}
\end{center}
\end{figure}

The main aim of the simulation is to investigate if any truncation of nonzero Schmidt coefficients
is possible. Figure \ref{figDJrank} shows the error in the computed value of ${\rm Prob}(0_00_10_20_3)$
(the discrepancy from zero) as a function of $m_{\rm trunc}$. The maximum Schmidt rank $m_{\rm max,max}$
observed during the simulation is also plotted in the figure as a function of $m_{\rm trunc}$.
It is clearly shown that $m_{\rm trunc}$ should be equal to or more than the exactly maximum Schmidt
rank that is observed in the absence of truncations; otherwise a significant numerical error appears.
(In Ref.~\cite{S12}, we have shown a similar result for a smaller circuit of the algorithm for a different
balanced function.)
As previously mentioned, the distribution of nonzero Schmidt coefficients is closely related
to this phenomenon. We show the distribution in Fig.~\ref{figDist}, which was taken at the
point where the second CNOT gate was being processed among the $2N_g+1$ CNOT gates in the middle
part of Fig.~\ref{figDJcirc}(a). It clearly depicts that none of the nonzero Schmidt coefficients
was negligible.

In addition, it was found that precision slightly more than the double precision was
required to perform a stable simulation even when we did not impose any truncation of nonzero
Schmidt coefficients, as shown in Table~\ref{tableDJprec}.
\begin{table}[pbt]\caption{\label{tableDJprec}Precision (prec) dependence
of the numerical error (Error) in the computed value of ${\rm Prob}(0_00_10_20_3)$.
No truncation of nonzero Schmidt coefficients was employed. The real time consumed for the
simulation is also shown (Time), which is an average over 10 trials of the same simulation
(the standard deviation is shown in the parentheses). The simulated circuit had 65 qubits because $N_g$ was set to 7.
Environment: Red Hat Enterprise Linux 6 64-bit OS,
Intel Xeon E7-8837 2.67GHz (2.80GHZ maximum) CPU, and 315GB RAM.}
\begin{center}
\begin{tabular}{|c|c|c|}\hline
prec & Error & Time [sec]\\\hline
$\le 55$ & $\begin{array}{l}\text{Convergence failure in}\\\text{some of eigenvectors}\end{array}$ & N/A\\
$56$ & $1.80\times 10^{-32}$ &  {341 \scriptsize (10.4)}\\
$57$ & $5.66\times 10^{-26}$ &  {335 \scriptsize (10.2)}\\
$58$ & $1.72\times 10^{-32}$ &  {338 \scriptsize (7.82)}\\
$59$ & $1.55\times 10^{-33}$ &  {379 \scriptsize (6.45)}\\
$60$ & $1.92\times 10^{-34}$ &  {348 \scriptsize (13.3)}\\
$\vdots$ & $\vdots$ & $\vdots$\\
$256$ & $5.91\times 10^{-139}$ &  {416 \scriptsize (8.50)}\\
$\vdots$ & $\vdots$ & $\vdots$\\
$512$ & $7.16\times 10^{-121}$ &  {709 \scriptsize (19.0)}\\
$\vdots$ & $\vdots$ & $\vdots$\\
$768$ & $1.78\times 10^{-460}$ &  {922 \scriptsize (19.7)}\\
$\vdots$ & $\vdots$ & $\vdots$\\
$1024$ & $6.50\times 10^{-615}$ &  {1620 \scriptsize (12.3)}\\
$\vdots$ & $\vdots$ & $\vdots$\\
$1280$ & $7.77\times 10^{-770}$ &  {2170 \scriptsize (16.9)}
\\\hline
\end{tabular}
\end{center}
\end{table}
For precision less than or equal to 55 bits, some of matrix diagonalization routines failed
after the half point of the circuit, which indicates an accumulated error due to precision
shortage. More specifically speaking about this case, a small non-Hermitian fraction due to
the accumulated error in a reduced density matrix resulted in the convergence error of
eigenvectors during an update of tensors.

\section{Discussion}\label{secDiscussion}
\subsection{Discussion on the ZKCM library}
With the ZKCM library, standard tasks in numerical matrix calculations
can be completed within the time consumption on the same order of magnitude
as that with the PARI library as we have seen in the comparison performed
in Sec.\ \ref{secPE}. The comparison also showed that ZKCM is slower than PARI,
even in a simple matrix multiplication, which suggests that basic arithmetic
operations make a certain gap in speed of the two libraries. In fact,
ZKCM uses MPFR functions while PARI uses GMP functions for the basic arithmetic
computation. MPFR is approximately two and a half times as slow as GMP in multiplying
floating-point numbers as far as we tested.\footnote{We wrote a test program to compute
$x\longleftarrow x * y$ with $y=1.00-1.00\times10^{-7}$ for $10^8$ times where the initial
value of $x$ was set to $1$. The precision was set to 512 bits. It took $15.3$
seconds on average (with standard deviation $1.09\times10^{-1}$) when we used MPFR's ``{\tt mpfr\_t}''
structure for floating point numbers while it took only $6.11$ seconds on average (with standard
deviation $4.91\times10^{-2}$) when we used GMP's ``{\tt mpf\_t}'' structure. The average was taken
over ten trials. This test was performed on a machine with the Fedora 15 64-bit OS, Intel Core
i5 M460 CPU, and 4GB RAM. GMP version 4.3.2 and MPFR version 3.0.0 were used.
} Thus it is reasonable to find a certain gap in computational speed.
We believe the gap is acceptable as long as it does not change the order of magnitude
of computation time for basic matrix computation.
It has been probably a historical reason that PARI has not used MPFR so far.\footnote{
PARI has been developed since 1979 \cite{PARI}. GMP appeared in 1991 \cite{GMP} and
MPFR appeared in 2000 \cite{MPFR}. PARI already had plenty of fast floating-point
functionalities at the time MPFR appeared.} 
We chose MPFR because it has floating-point exceptions and other useful functionalities
for real floating-point numbers by default.

The only unexpected result in Sec.\ \ref{secPE} is the one about Hermitian
matrix diagonalization shown in Tables \ref{tableHerm} and \ref{tableHerm2}.
It has been shown that time consumptions of functions ``\verb|diag_H|'' of ZKCM
and ``\verb|eigen|'' and ``\verb|jacobi|'' of PARI are on the same order of magnitude.
This is an unusual result because the Jacobi method employed in ``\verb|jacobi|''
looks as fast as a variant of the QR method employed in ``\verb|diag_H|'' for
a $100\times100$ matrix.

It is likely that ``\verb|diag_H|'' and ``\verb|jacobi|'' are used for a similar purpose
or under a similar situation, since they both work fine for a matrix with degenerate
eigenvalues. (It is known \cite{pariknown} that ``\verb|eigen|'' almost always fails
for a matrix with degenerate eigenvalues.) In this point of view, too, it is meaningful
to compare ``\verb|diag_H|'' and ``\verb|jacobi|''.

As mentioned above, function ``\verb|jacobi|'' of PARI is an implementation of a
standard Jacobi method. Function ``\verb|diag_H|'' of ZKCM employs the Householder-QR method
with the Wilkinson shift for computing approximate eigenvalues and uses several sets
of inverse iterations to find eigenvectors and at the same time to enhance the accuracy of
eigenvalues.
It has been commonly known that the Jacobi method is slower than the QR method:
 ``For matrices of order greater than about 10, however, the algorithm is slower, by
a significant constant factor, than the QR method...''|page 571 of Ref.~\cite{Nrecipe};
``...the Jacobi method is several times slower than a reduction to tridiagonal form,
followed by Francis's algorithm.''|page 488 of Ref.~\cite{Wat10}.
Thus, it is unusual that, for a relatively large matrix with the order $100$, there
is no significant advantage in using the Householder-QR method. In a phenomenological
explanation, this is due to the large cost of inverse iterations.
By using the Gprof program \cite{gprof} (a monitoring software), it turned out that,
in ``\verb|diag_H|'', more than 71.1 percent of running time was spent for the inverse
iterations in case of a $100\times100$ matrix with the precision $\ge 512$ [bits]. In fact
speedup of one or two orders of magnitude was possible by omitting inverse iterations
for finding eigenvalues as we have shown in Tables \ref{tableEI1} and \ref{tableEI2}.
It was however impossible to achieve a required precision without inverse iterations
as clearly shown in the tables.

Thus, it was unexpectedly expensive to achieve a sufficient convergence by inverse iterations.
We, of course, used the LU decomposition to reduce the cost of each inverse iteration. This, however,
did not mitigate the total cost of the inverse iterations sufficiently. As a matter of fact, the
total number of the inverse iterations had to be increased along with the increase in required
precision. This, at a glance, does not look in accordance with a conventional theory of inverse
iteration \cite{I97} suggesting that the machine epsilon does not make a difference in the process of
inverse iteration. The conventional theory, however, considers the process where the inverse iteration
is used for computing an eigenvector for a given approximate eigenvalue. In our case, in contrast, we
also improve accuracy of the eigenvalue using the computed eigenvector. A routine of inverse iteration
is often called several times in order for achieving sufficient accuracy for each pair of an
eigenvalue and the corresponding eigenvector. Furthermore, program routines called by the routine
of inverse iteration are not error-free in practice. Thus it was not surprising that the choice of
precision considerably affected the cost of inverse iterations (and hence the time consumption of
``\verb|diag_H|'' shown in Table \ref{tableEI1}).

Another factor that makes inverse iterations expensive is the cost of basic arithmetic
operations in multiprecision computing. Unlike double-precision computing where every
basic arithmetic instruction is performed within a few clock cycles, it takes quite
many cycles.\footnote{Theoretically, each instruction with precision ${\rm prec}$ takes
$O({\rm prec}^c)$ time with $c\le 2$ dependent on the chosen algorithm. In real CPU time, it
should scale slightly better in practice because of vectorization of operands. See the
manual of GMP \cite{GMP} for the details of algorithms for arithmetic instructions.}
It is fast in double-precision computing to successively perform arithmetic instructions
because this does not involve any conditional branching. In contrast, it is inevitable to
involve several conditional branching instructions in every arithmetic operation in multiprecision
computing. Thus, the simple fact that basic arithmetic instructions are not hardware instructions
should be a large factor.

In future, we should find a strategy for faster Hermitian matrix diagonalization. As a possible
approach, the precision of computation for finding approximate eigenvalues can be internally enlarged
so that a relatively small number of subsequent inverse iterations should be enough to achieve a
required precision. Table \ref{tableEI1}, however, shows that accuracy of approximate eigenvalues is
not very much enhanced by simply increasing the preset precision in the Householder-QR method. So far,
the author could not find the reason of this tendency. A detailed theoretical and practical analyses of
the present implementation (and perhaps the method itself) will be required to overcome this difficulty.
For another approach, there is a possibility that a certain method uncommon under the double-precision
environment possesses an advantage under a high-precision environment in a practical sense. It is of
interest to re-investigate practical usefulness of conventional methods \cite{Saad11} for eigenvalue
problems under a high-precision environment.

Apart from computational speed, the syntax of a library is also important for usability.
As a C++ library, ZKCM provides an easy-to-use syntax for handling matrices by operator
overloading. It also provides various functionalities for matrix computation
as member functions of a class as well as as external functions. Thus a program
code using the library should look simpler than those written with some other
multiprecision libraries for Fortran or C languages. In general, it is relatively
easy to write an extension library of a C++ library. As we have introduced, ZKCM has an
extension library named ZKCM\_QC developed for a (time-dependent) MPS simulation of
quantum circuits. The main library and the extension are under steady development.
Latest development versions can be downloaded from the repositories linked from the
URL~\cite{ZKCM}.

\subsection{Discussion on the ZKCM\_QC library}
As for the ZKCM\_QC library, it is worth mentioning that multiprecision computation is
useful in MPS simulations of quantum computing as discussed in Sec.~\ref{secDJ}. Indeed, truncation
errors are dominant whenever truncations of nonzero Schmidt coefficients are employed. It is however
uncommon to employ such truncations in simulating quantum computing since they cause an unacceptably large
error as we have discussed. As a consequence, the rounding error become the only possible error.
In our example, slightly more than double precision was required for reliable MPS simulation. This
is true in a different example to simulate quantum search, which is shown in our related
contribution \cite{S12}. Thus, time-dependent MPS simulation is fragile not only against
accumulating truncation errors \cite{Gob05,JJGR06} but also against accumulating rounding errors.
High-precision computation is therefore beneficial to the MPS method.

Apart from the MPS method, there have been several simulators of quantum computing which make
use of parallel programming techniques \cite{Niwa02,R07,Ta09,Gu10} to mitigate the time consumption
of the brute-force method. They use a parallelized exact matrix computation and spend a massive
computational resource ({\em e.g.}, more than one thousand CPU processes and several hundred gigabytes
physical memory to handle less than forty qubits \cite{R07}).
It seems that an accumulation of rounding errors is not significant for these simulators although they
use double precision computation, as this was not reported so far. In contrast to their approach, use
of the MPS method is quite economical. As we have shown in Sec.~\ref{secDJ}, we could handle 65 qubits in
the MPS simulation of the Deutsch-Jozsa algorithm under a certain setup, which ran as a single CPU
process\footnote{Every simulation with ZKCM or ZKCM\_QC shown in this paper was run as a single CPU process.}
and took only (approximately) seven minutes in case of 256-bit floating-point precision.

Besides the MPS method, Viamontes {\em et al.}'s method \cite{VMH03,VMH05,Via07} using a compressed
graph representation is also quite economical to simulate quantum computation. According to
Refs.~\cite{VMH03,Via07}, its compressed data structure is sensitive to rounding errors so that they used
multiprecision computation based on the GMP library. The MPS data structure (\ref{eq1}) is of course
a compressed data structure for which we needed multiprecision computation for stable quantum circuit
simulation. It will be interesting to theoretically investigate if a simulation of quantum computing
utilizing a compressed data structure generally has a certain inevitable sensitivity to rounding errors.

Finally, we discuss on the computational cost of the MPS method, which is known to grow polynomially in
the maximum Schmidt rank $m_{\rm max,max}$ during the simulation \cite{V03} as mentioned in Sec.~\ref{sectheory}.
It is expected that MPS simulation becomes very expensive for quantum circuits of algorithms for hard
problems like those for quantum prime factorization \cite{Shor} of a large composite number. It has been
discussed \cite{JL03} that large entanglement (this usually leads to a large Schmidt rank) must be involved
in quantum prime factorization. So far, Kawaguchi {\em et al.} \cite{KQIT11} found Schmidt rank $92$ in a
modular exponentiation circuit (this is used in quantum prime factorization) with $35$ qubits in their MPS
simulation. Nevertheless, it has been neither proved nor numerically verified that $m_{\rm max,max}$ grows
exponentially in the number $n$ of qubits as far as the author knows.
Although there have been several studies on entanglement during quantum prime factorization
\cite{PP02,SSB05,KM06,MSB10}, they have not reached an answer to how entanglement grows in $n$.
It is still an open problem how we rigorously estimate the value of $m_{\rm max,max}$ for a given quantum circuit.
Presently, a known upper bound for $m_{\rm max,max}$ is a function exponentially growing in the number of basic
quantum gates overlapping to each other ({\em i.e.}, those stretching across the same bundle of wires) \cite{J06};
this is however not practically useful for a user of the MPS method to estimate the value of $m_{\rm max,max}$
for his/her simulation. More theoretical and numerical efforts are required to understand the scalability of
the method.

\section{Summary}\label{secSummary}
We have introduced the ZKCM library which is a C++ library developed for multiprecision
complex-number matrix computation. It is especially usable for high-precision numerical
simulations in quantum physics; it has an easy-to-use syntax for matrix manipulations and
helpful functionalities like the tensor-product and tracing-out operations, the discrete Fourier
transform, etc. An extension library ZKCM\_QC has also been introduced, which is a library
for a multiprecision time-dependent matrix-product-state simulation of quantum computing.
It enables a user-friendly coding for simulating quantum circuits.

\appendix
\section{Updating tensors of an MPS for simulating a three-qubit gate operation}\label{appendix}
The library ZKCM\_QC uses three-qubit gates as elementary gates in addition to
single- and two-qubit gates. As it is not usually explained in detail how a three-qubit
gate is simulated in a time-dependent MPS simulation, a detailed explanation is
given here. As for simulations of single- and two-qubit gates, see Refs.\ \cite{V03,S06}
for detailed explanations.

Consider the quantum gate
\[
\begin{split}
U =&\sum_{i_li_{l+1}i_{l+2}=000}^{111}
\sum_{k_lk_{l+1}k_{l+2}=000}^{111}
u_{i_li_{l+1}i_{l+2},k_lk_{l+1}k_{l+2}}\\
&\times|i_li_{l+1}i_{l+2}\rangle
\langle k_lk_{l+1}k_{l+2}|
\end{split}
\]
acting on the three qubits $l$, $l+1$, and $l+2$.
With a focus on the three qubits, the MPS of $n$ qubits can be written as
\[
\begin{split}
&|\Psi\rangle=\sum_{i_li_{l+1}i_{l+2}=000}^{111}
\sum_{v_{l-1},v_{l},v_{l+1},v_{l+2}=0,0,0,0}^{m_{l-1}-1,m_{l}-1,m_{l+1}-1,m_{l+2}-1}\biggl[
Q_{l}(i_l,v_{l-1},v_l)\\
&\times V_l(v_l)Q_{l+1}(i_{l+1},v_l,v_{l+1})
V_{l+1}(v_{l+1})Q_{l+2}(i_{l+2},v_{l+1},v_{l+2})\\
&\times |v_{l-1}^{0,\ldots,l-1}\rangle|i_l\rangle|i_{l+1}\rangle|i_{l+2}\rangle
|v_{l+2}^{l+3,\ldots,n-1}\rangle\biggr]
\end{split}
\]
with $|v_{l-1}^{0,\ldots,l-1}\rangle=V_{l-1}(v_{l-1})|\Phi_{v_{l-1}}^{0,\ldots,l-1}\rangle$
and $|v_{l+2}^{l+3,\ldots,n-1}\rangle=V_{l+2}(v_{l+2})|\Phi_{v_{l+2}}^{l+3,\ldots,n-1}\rangle$,
where $|\Phi_{v_{l-1}}^{0,\ldots,l-1}\rangle$ are
the Schmidt vectors of the block of qubits $0,\ldots, l-1$ for the splitting between
$l-1$ and $l$, and $|\Phi_{v_{l+2}}^{l+3,\ldots,n-1}\rangle$ are those of the block of
qubits $l+3,\ldots,n-1$ for the splitting between $l+2$ and $l+3$.

What we should do as a simulation of applying the quantum gate $U$ to $|\Psi\rangle$
is to update tensors 
${Q_{l}}$,
${V_l}$,
${Q_{l+1}}$,
${V_{l+1}}$, and
${Q_{l+2}}$ to 
$\widetilde{Q_{l}}$,
$\widetilde{V_l}$,
$\widetilde{Q_{l+1}}$,
$\widetilde{V_{l+1}}$, and
$\widetilde{Q_{l+2}}$, respectively, so that the MPS with
the updated tensors represents the resultant state $|\widetilde\Psi\rangle$.
This process is explained hereafter step by step. (Here is some note on the description:
In case $l-1=-1$, the tensor $V_{l-1}$ should be regarded as unity and the parameter $v_{l-1}$
should be dropped. Similarly, in case $l+2=n-1$, the tensor $V_{l+2}$ should be regarded as unity
and the parameter $v_{l+2}$ should be dropped.)

The state after $U$ is applied to the qubits $l$, $l+1$, and $l+2$ can be written as
\begin{equation}\label{resultant}
\begin{split}
|\widetilde\Psi\rangle=&\sum_{v_{l-1}}\sum_{v_{l+2}}
\sum_{i_li_{l+1}i_{l+2}}
\Theta(i_l,i_{l+1},i_{l+2},v_{l-1},v_{l+2})\\
&\times |v_{l-1}^{0,\ldots,l-1}\rangle|i_l\rangle|i_{l+1}\rangle|i_{l+2}\rangle
|v_{l+2}^{l+3,\ldots,n-1}\rangle
\end{split}
\end{equation}
with the tensor
\[\begin{split}
&\Theta(i_l,i_{l+1},i_{l+2},v_{l-1},v_{l+2})
=\sum_{v_l}\sum_{v_{l+1}}\sum_{k_lk_{l+1}k_{l+2}}\biggl[\\
&~~~~u_{i_li_{l+1}i_{l+2},k_lk_{l+1}k_{l+2}}
Q_{l}(k_l,v_{l-1},v_l)V_l(v_l)\\
&\times Q_{l+1}(k_{l+1},v_l,v_{l+1})
V_{l+1}(v_{l+1})Q_{l+2}(k_{l+2},v_{l+1},v_{l+2})\biggr].
\end{split}\]

First, we are going to compute the tensors
$\widetilde{Q_{l}}(i_l,v_{l-1},v_l)$ and $\widetilde{V_l}(v_l)$
of the resultant state.
The reduced density matrix of qubits $0,...,l$ calculated from
Eq.\ (\ref{resultant}) is
\[\begin{split}
&\rho^{0,\ldots,l} = \sum_{i_l v_{l-1} {i'}_l {v'}_{l-1}}\biggl[
\sum_{i_{l+1}i_{l+2}v_{l+2}}
[V_{l+2}(v_{l+2})]^2\\
&\times\Theta(i_l,i_{l+1},i_{l+2},v_{l-1},v_{l+2})
\Theta^*({i'}_l,i_{l+1},i_{l+2},{v'}_{l-1},v_{l+2})\biggr]\\
&\times|v_{l-1}\rangle|i_l\rangle\langle {v'}_{l-1}|\langle {i'}_l|\\
&=
\sum_{i_l v_{l-1} {i'}_l {v'}_{l-1}}\biggl[
\sum_{i_{l+1}i_{l+2}v_{l+2}}
[V_{l+2}(v_{l+2})]^2\\
&\times \Theta(i_l,i_{l+1},i_{l+2},v_{l-1},v_{l+2})
\Theta^*({i'}_l,i_{l+1},i_{l+2},{v'}_{l-1},v_{l+2})\\
&\times V_{l-1}(v_{l-1})V_{l-1}({v'}_{l-1})\biggr]
|\Phi_{v_{l-1}}\rangle|i_l\rangle
\langle \Phi_{{v'}_{l-1}}|\langle {i'}_l|\\
&=\sum_{i_l v_{l-1} {i'}_l {v'}_{l-1}}\hspace{-2.2mm}
a_{i_l v_{l-1} {i'}_l {v'}_{l-1}}
|\Phi_{v_{l-1}}\rangle|i_l\rangle
\langle \Phi_{{v'}_{l-1}}|\langle {i'}_l|
\end{split}\]
with $a_{i_l v_{l-1} {i'}_l {v'}_{l-1}}=\left[
\sum_{i_{l+1}i_{l+2}v_{l+2}}\cdots\right]$.
The matrix $\rho^{0,\ldots,l}$ is a $(2m_{l-1})\times(2m_{l-1})$
matrix. This is now diagonalized to achieve
the eigenvalues $\widetilde{\lambda}_{v_l} =[\widetilde{V_{l}}(v_{l})]^2$
and the corresponding eigenvectors $|\widetilde{\Phi}_{v_l}^{0,\ldots,l}\rangle$
under the basis $\{|\Phi_{v_{l-1}}^{0,\ldots,l-1}\rangle|i_l\rangle\}$.
Immediately we find the values of $\widetilde{V_{l}}(v_{l})$.
We may truncate out negligibly small eigenvalues\footnote{We should not employ a
threshold for the number of eigenvalues as we have discussed in Sec.~\ref{secDJ}.}
to reduce $\widetilde{m_l}$ (the updated value of $m_l$, namely, the number of data
in $\widetilde{V_{l}}$).
In addition, we have vector elements $C_l(i_l,v_{l-1},v_l)$ of just
computed eigenvectors:
\[
|\widetilde{\Phi}_{v_l}^{0,\ldots,l}\rangle = C_l(i_l,v_{l-1},v_l)
|\Phi_{v_{l-1}}^{0,\ldots,l-1}\rangle|i_l\rangle,
\]
from which we can derive
\[
\widetilde{Q_l}(i_l,v_{l-1},v_l) = C_l(i_l,v_{l-1},v_l)/V_{l-1}(v_{l-1}).
\]
Thus we have obtained
\[
|\widetilde{\Phi}_{v_l}^{0,\ldots,l}\rangle
=\widetilde{Q_l}(i_l,v_{l-1},v_l)|v_{l-1}\rangle|i_l\rangle.
\]

Second, we are going to compute the tensors
$\widetilde{V_{l+1}}(v_{l+1})$ and
$\widetilde{Q_{l+2}}(i_{l+2},v_{l+1},v_{l+2})$ from Eq. (\ref{resultant}).
The reduced density matrix of qubits $l+2,\ldots,n-1$ is
\[\begin{split}
&\rho^{l+2,\ldots,n-1}=\sum_{i_{l+2}v_{l+2}{i'}_{l+2}{v'}_{l+2}}\biggl[
\sum_{i_li_{l+1}v_{l-1}}[V_{l-1}(v_{l-1})]^2\\
&\times\Theta(i_l,i_{l+1},i_{l+2},v_{l-1},v_{l+2})
\Theta^*(i_l,i_{l+1},{i'}_{l+2},v_{l-1},{v'}_{l+2})
\biggr]\\
&\times|i_{l+2}\rangle|v_{l+2}\rangle\langle{i'}_{l+2}|\langle{v'}_{l+2}|\\
&=\sum_{i_{l+2}v_{l+2}{i'}_{l+2}{v'}_{l+2}}\biggl[
\sum_{i_li_{l+1}v_{l-1}}[V_{l-1}(v_{l-1})]^2\\
&\times\Theta(i_l,i_{l+1},i_{l+2},v_{l-1},v_{l+2})
\Theta^*(i_l,i_{l+1},{i'}_{l+2},v_{l-1},{v'}_{l+2})\\
&\times V_{l+2}(v_{l+2})V_{l+2}({v'}_{l+2})\biggr]
|i_{l+2}\rangle|\Phi_{v_{l+2}}\rangle
\langle{i'}_{l+2}|\langle\Phi_{{v'}_{l+2}}|\\
&=\sum_{i_{l+2}v_{l+2}{i'}_{l+2}{v'}_{l+2}}
{~\hspace{-9mm}~}b_{i_{l+2}v_{l+2}{i'}_{l+2}{v'}_{l+2}}
|i_{l+2}\rangle|\Phi_{v_{l+2}}\rangle
\langle{i'}_{l+2}|\langle\Phi_{{v'}_{l+2}}|
\end{split}
\]
with $b_{i_{l+2}v_{l+2}{i'}_{l+2}{v'}_{l+2}}=
\left[\sum_{i_li_{l+1}v_{l-1}}\cdots\right]$.
The matrix $\rho^{l+2,\ldots,n-1}$ is a $(2m_{l+2})\times(2m_{l+2})$
matrix. This is now diagonalized to achieve the eigenvalues
$\widetilde{\lambda}_{v_{l+1}}=[\widetilde{V_{l+1}}(v_{l+1})]^2$
and the corresponding eigenvectors $|\widetilde{\Phi}_{v_{l+1}}^{l+2,\ldots,n-1}\rangle$
under the basis $\{|i_{l+2}\rangle|\Phi_{v_{l+2}}\rangle\}$.
The values of $\widetilde{V_{l+1}}(v_{l+1})$ are immediately found.
We may truncate out negligibly small eigenvalues to reduce $\widetilde{m_{l+1}}$
(the updated value of $m_{l+1}$).
In addition, we have vector elements $C_{l+2}(i_{l+2},v_{l+1},v_{l+2})$ of
just computed eigenvectors:
\[
|\widetilde{\Phi}_{v_{l+1}}^{l+2,\ldots,n-1}\rangle = C_{l+2}(i_{l+2},v_{l+1},v_{l+2})
|i_{l+2}\rangle|\Phi_{v_{l+2}}^{l+3,\ldots,n-1}\rangle,
\]
from which we can derive
\[
\widetilde{Q_{l+2}}(i_{l+2},v_{l+1},v_{l+2}) =
C_{l+2}(i_{l+2},v_{l+1},v_{l+2})/V_{l+2}(v_{l+2}).
\]
Thus we have obtained
\[
|\widetilde{\Phi}_{v_{l+1}}^{l+2,\ldots,n-1}\rangle
=\widetilde{Q_{l+2}}(i_{l+2},v_{l+1},v_{l+2})|i_{l+2}\rangle|v_{l+2}\rangle.
\]

Third, we are going to compute the tensor
$\widetilde{Q_{l+1}}(i_{l+1},v_{l},v_{l+1})$.
By the definition of this tensor, we have
\[\begin{split}
&\widetilde{Q_{l+1}}(i_{l+1},v_{l},v_{l+1})
=\frac{
\langle\widetilde{\Phi}_{v_l}^{0,\ldots,l}|
\langle i_{l+1}|\langle\widetilde{\Phi}_{v_{l+1}}^{l+2,\ldots,n-1}|
\widetilde\Psi\rangle}
{\widetilde{V_l}(v_l)\widetilde{V_{l+1}}(v_{l+1})}\\
&=\frac{1}{\widetilde{V_l}(v_l)\widetilde{V_{l+1}}(v_{l+1})}
\sum_{i_li_{l+2}v_{l-1}v_{l+2}}\biggl\{
\left(\langle\widetilde{\Phi}_{v_l}^{0,\ldots,l}|v_{l-1}\rangle|i_l\rangle\right)\\
&\times\left(\langle\widetilde{\Phi}_{v_{l+1}}^{l+2,\ldots,n-1}
|i_{l+2}\rangle|v_{l+2}\rangle\right)
\Theta(i_l,i_{l+1},i_{l+2},v_{l-1},v_{l+2})\biggr\}.
\end{split}\]
We substitute the equations
\[
 \left\{
\begin{array}{l}
|\widetilde{\Phi}_{v_l}^{0,\ldots,l}\rangle
=\widetilde{Q_{l}}(i_l,v_{l-1},v_l)|v_{l-1}\rangle|i_l\rangle\\
|\widetilde{\Phi}_{v_{l+1}}^{l+2,\ldots,n-1}\rangle
=\widetilde{Q_{l+2}}(i_{l+2},v_{l+1},v_{l+2})|i_{l+2}\rangle|v_{l+2}\rangle
\end{array}
\right.
\]
into the above equation to obtain
\[\begin{split}
&\widetilde{Q_{l+1}}(i_{l+1},v_l,v_{l+1})=
\frac{1}{\widetilde{V_l}(v_l)\widetilde{V_{l+1}}(v_{l+1})}
\sum_{i_li_{l+2}v_{l-1}v_{l+2}}\biggl\{\\
&[V_{l-1}(v_{l-1})]^2\widetilde{Q_{l}}^*(i_l,v_{l-1},v_l)
\widetilde{Q_{l+2}}^*(i_{l+2},v_{l+1},v_{l+2})\\
&\times [V_{l+2}(v_{l+2})]^2
\Theta(i_l,i_{l+1},i_{l+2},v_{l-1},v_{l+2})
\biggr\}.
\end{split}\]

In this way, we have obtained the updated tensors
\[\begin{split}
&\widetilde{Q_{l}}(i_l,v_{l-1},v_l),~
\widetilde{V_l}(v_l),~
\widetilde{Q_{l+1}}(i_{l+1},v_l,v_{l+1}),\\
&\widetilde{V_{l+1}}(v_{l+1}),~{\rm and}~
\widetilde{Q_{l+2}}(i_{l+2},v_{l+1},v_{l+2}).
\end{split}\]

The described process to simulate a three-qubit gate is implemented
as the function ``\verb|applyU8|'' in ZKCM\_QC. We will see an example
to use the function in the following.
\subsection{Example}\label{simplestGrover}
Here is an example program to use a three-qubit operation.
It simulates the simplest case of Grover's quantum search \cite{Gr96}.
One may also see the simulation performed in Sec.~\ref{secDJ} for another example.

For a brief sketch of the quantum search, suppose that there is an
oracle function $f:\{0,1\}^n\rightarrow\{0,1\}$
that has $r$ solutions ${\bf w}$ (namely, strings ${\bf w}$ such that $f({\bf w})=1$).
Classical search for finding a solution takes $O(2^n/r)$ queries. In contrast,
the Grover search finds a solution in $O(\sqrt{2^n/r})$ queries.
In particular when $n=4$ and $r=1$, a single query is enough for the
Grover search. For example, consider the parent data set $\{00, 01, 10, 11\}$
and the solution $01$ for a certain oracle with a single oracle bit.
A single Grover iteration $R$ maps $|s\rangle=\frac{1}{2}(|00\rangle
+|01\rangle+|10\rangle+|11\rangle)|-\rangle_{\rm o}$ to
$|01\rangle|-\rangle_{\rm o}$,
where $R=U_sU_f$ with $U_f=1-2|01\rangle\langle01|$ and
$U_s=1-2|s\rangle\langle s|$ acts on the left side qubits;
$|-\rangle_{\rm o}=(|0\rangle_{\rm o}-|1\rangle_{\rm o})/\sqrt{2}$ is a state
of the oracle qubit (the right-most qubit). $U_f$ is a unitary operation
corresponding to $f$ and $U_s$ is a so-called inversion-about-average operation.
To implement $U_f$ and $U_s$, we utilize the fact that flipping
$|-\rangle_{\rm o}$ changes the phase factor $\pm 1$, following a common
implementation \cite{NC2000}.

Here, we consider a very simple oracle structure, namely that, it
flips $|-\rangle_{\rm o}$ when the left side qubits are in $|01\rangle$
using a very straight-forward circuit interpretation. Note that there is
no realistic benefit to perform a search in such a case because we know
the solution beforehand. Usually a search is performed because we cannot
guess the solutions by the circuit appearance (consider, {\em e.g}, an instance
of a satisfiability problem \cite{GJ79}). The sample program to simulate the
Grover search for the present simple case is shown in Listing~\ref{codeqsearch}.
\begin{lstlisting}[label=codeqsearch,caption=simple\_q\_search.cpp]
#include "zkcm_qc.hpp"
int main()
{
  zkcm_set_default_prec(256);
  mps M(3);
  //Prepare the state for the parent data set and
  //the initial state of the oracle qubit.
  M.applyU(tensor2tools::Hadamard, 0);
  M.applyU(tensor2tools::Hadamard, 1);
  M.applyU(tensor2tools::PauliX, 2);
  M.applyU(tensor2tools::Hadamard, 2);

  //Oracle function with the solution 01.
  zkcm_matrix U_f(8, 8);
  U_f.set_to_identity();
  U_f(2, 2) = U_f(3, 3) = 0;
  U_f(2, 3) = U_f(3, 2) = 1;

  //Inversion-about-average operation
  zkcm_matrix U_s(8, 8);
  U_s.set_to_identity();
  U_s(0, 0) = U_s(1, 1) = 0;
  U_s(0, 1) = U_s(1, 0) = 1;
  zkcm_matrix H = zkcm_matrix("[1, 1; 1, -1]")
                  / sqrt(zkcm_class(2));
  zkcm_matrix I = zkcm_matrix("[1, 0; 0, 1]");
  H = tensorprod(tensorprod(H, H), I);
  U_s = H * U_s * H;//This is 1-2|s><s|.
 
  std::cout << "Initially, Prob(01)="
            << M.RDO_block(0, 1).get(1, 1)
            << std::endl
            << "Go into Grover's iteration..."
            << std::endl;
  for (int i = 0; i < 8; i++)
  {
    M.applyU8(U_f, 0, 1, 2);
    M.applyU8(U_s, 0, 1, 2);
    std::cout << "After " << i + 1
              << " times iteration, Prob(01)="
              << M.RDO_block(0, 1).get(1, 1)
              << std::endl;
  }
  zkcm_quit();
  return 0;
}
\end{lstlisting}
The program is compiled and executed in the following way.
\begin{Verbatim}
[user@localhost foo]$ c++ -o simple_q_search \
simple_q_search.cpp -lzkcm_qc -lzkcm -lmpfr \
-lgmp -lgmpxx
[user@localhost foo]$ ./simple_q_search 
Initially, Prob(01)=2.500000000e-01
Go into Grover's iteration...
After 1 times iteration, Prob(01)=1.000000000e+00
After 2 times iteration, Prob(01)=2.500000000e-01
After 3 times iteration, Prob(01)=2.500000000e-01
After 4 times iteration, Prob(01)=1.000000000e+00
After 5 times iteration, Prob(01)=2.500000000e-01
After 6 times iteration, Prob(01)=2.500000000e-01
After 7 times iteration, Prob(01)=1.000000000e+00
After 8 times iteration, Prob(01)=2.500000000e-01
\end{Verbatim}
We see that the success probability, {\em i.e.}, the probability to find
$01$ in the left side qubits if we measure them in the computational
basis, is unity after a single $R$ is applied.
By continuing the iteration, the success probability oscillates. 

We have seen a sample program that uses a three-qubit gate by calling the
function ``\verb|applyU8|''. In addition to this function, there are
functions for typical three-qubit gates. They are briefly mentioned below.
\subsection{Typical three-qubit gates}
The first typical three-qubit gate is the CCNOT gate already introduced in Sec.~\ref{secDJ},
which flips a single target qubit $t$ under the condition that two control qubits
$a$ and $b$ are in $|11\rangle$. ZKCM\_QC has a particular member function (of class
``\verb|mps|'') for it:
``\verb|mps & mps::CCNOT (int a, int b, int t);|''.
The second typical one is the controlled-SWAP (CSWAP) gate,
which swaps two target qubits $p$ and $q$ under the condition that the control qubit $c$
is in $|1\rangle$. The member function corresponding to this gate is
``\verb|mps & mps::CSWAP (int c, int p, int q);|''.

In summary of this appendix, the process to simulate a three-qubit
gate operation in a time-dependent MPS simulation has been theoretically
described. A sample program to simulate a simple quantum search has been
provided, which uses one of the three-qubit gate functions of the ZKCM\_QC
library.
\renewcommand{\refname}{References}

\end{document}